\documentclass{jfm}
\usepackage{lineno,hyperref}
\usepackage{graphicx}
\usepackage{floatrow}
\usepackage{epstopdf, epsfig}
\usepackage{upgreek}
\usepackage{subfigure}
\usepackage{amsmath}
\usepackage{amssymb}
\usepackage{mathrsfs}  
\usepackage{nomencl}
\usepackage{setspace}
\usepackage[round]{natbib}
\usepackage[usenames,dvipsnames]{xcolor}
\usepackage{tikz}
\usetikzlibrary{shapes}
\usepackage{MnSymbol}
\usepackage[normalem]{ulem}
\usepackage[ruled,vlined]{algorithm2e}
\usepackage{wrapfig}
\usepackage{colortbl}
\usepackage{mathtools}

\shorttitle{Variational data assimilation in wall turbulence}
\shortauthor{M. Wang and T. A. Zaki}

\def\m1line{\vrule width3pt height2.5pt depth -2pt}

\def\bdot{\raise.2em\hbox to .15em{.}}

\usepackage{todonotes}
\title{\Large Variational data assimilation in wall turbulence: \\ 
From outer observations to wall stress and pressure}

\author{
Mengze Wang
\and  Tamer A.~Zaki
\corresp{\email{t.zaki@jhu.edu}}
}

\affiliation{Department of Mechanical Engineering, 
Johns Hopkins University,
Baltimore, MD 21218, USA}

\begin{document}

\maketitle

\begin{abstract}
Estimation of near-wall turbulence in channel flow from outer observations is investigated using adjoint-variational data assimilation.
We first consider fully resolved velocity data, starting at a distance from the wall.
By enforcing the estimated flow to exactly satisfy the Navier-Stokes equations, we seek a statistically stationary turbulent state that reproduces the instantaneous outer measurements.
Such an estimated state provides full access to the unknown near-wall turbulence, including the wall shear stresses and pressure.
When the first observation is within fifty viscous units from the wall, the correlation coefficient between the true and estimated state exceeds 95\%.
As the observations are further separated from the wall, at ninety viscous units, 
the accuracy of the assimilated wall stresses decreases to 40\% at the wall.
This trend is nearly independent of the Reynolds number.
The Fourier spectrum of the estimation error is qualitatively consistent with the coherence spectrum between the outer and the inner state variables: observed long wavelength structures in the outer flow have deeper coherence into the unobserved near-wall region, and therefore the error is lowest at large scales.
Nevertheless, the adjoint-variational approach provides a more rigorous quantification of the capacity to accurately predict the instantaneous near-wall turbulence from outer measurements.  Lastly, we demonstrate the robustness of the estimation accuracy using filtered and sub-sampled outer observations.

\end{abstract}

\section{Introduction}

Near-wall turbulence and the associated surface stresses are challenging to directly probe in experiments
\citep{Hutchins2009hotwire,Lee2016PIV}.
The near-wall region is, however, instrumental to understanding the dynamics of wall turbulence, such as the dissipation of energy, generation of drag, and extreme stress events.
In order to estimate the inner-layer turbulence from outer measurements, various reduced-complexity models have been proposed \citep{Marusic2010science,Illingworth2018,Sasaki2019}.
A recent study utilizing the Navier-Stokes equations demonstrated  that turbulence in the near-wall region can be reconstructed with machine accuracy, when fully resolved measurements in the outer flow are imposed at every time step. This process is termed synchronization because the inner layer synchronizes to the true flow that generated the outer observations, and is only possible if the first observation point is within approximately thirty viscous units from the wall.
Beyond this critical thickness synchronization fails, and the highest accuracy of estimating the wall layer under such conditions has never been examined.
This problem is investigated herein using four-dimensional adjoint-variational (4DVar) data assimilation. We report the optimal reconstruction of all the missing scales of near-wall turbulence, including the wall shear stresses and pressure, and the dependence of estimation accuracy on the available observations.  

The methodology adopted here is from the field of data assimilation, and the focus is the estimation of near-wall turbulence from under-resolved outer measurements. Such estimation is, however, relevant in other scenarios, e.g.\,wall modeling where the stress at the wall is estimated from outer velocities in large-eddy simulations.  We recall some of those approaches, although we do not provide an exhaustive review, and simply distinguish them from the variational data-assimilation method.
The earliest efforts to estimate wall stresses from outer observations are based on the law of the wall.  
By assuming a logarithmic profile for the instantaneous streamwise velocity in the outer layer, the wall shear stress at the same horizontal location as the velocity can be directly evaluated \citep{Schumann1975eqwm,Grotzbach1987eqwm}.
This model was further improved by introducing a streamwise separation between the available velocity data and the predicted wall shear stress \citep{Piomelli1989shift}, to account for the existence of streamwise-inclined coherent structures that are responsible for the fluctuations of both the outer velocity and the wall stress \citep{Rajagopalan1979inclined}. 
With advancements in our understanding of turbulent motions in the wall layer, alternative approaches have been developed to estimate near-wall turbulence, such as the ejection model that accounts for the correlation between vertical motions and wall stress \citep{Piomelli2008wmles}, and the inner-outer interaction model inspired by the amplitude modulation mechanism \citep{Marusic2010science,Baars2016}.  
In addition to physics-based models, data-driven methods have also been developed to predict wall stress using outer observations
\citep[e.g.][]{Bae2022RL}.
Nevertheless, all these models operate on a single-input-single-output basis, using velocity data from a single point in space and at the local time to estimate the wall stress at a corresponding spatial location and at the same time.  This property restricts the estimation accuracy, especially when multiple observations are available. 
In addition, reconstruction of pressure from a single observation is difficult due to the elliptic property of pressure in incompressible flows.

By incorporating multiple observations simultaneously from the outer layer, more flow features in the wall layer should be reconstructed than single-observation estimators.
\cite{Illingworth2018} considered fully resolved velocity observations in a horizontal plane, which are fed into a linear Navier-Stokes-based model augmented with eddy viscosity.
The large-scale structures below the observation plane were reconstructed with a reasonable accuracy.
\cite{Sasaki2019} proposed an estimator based on a data-driven transfer function, and demonstrated that the performance of the estimator can be improved by including multiple observation planes and quadratic nonlinearity.
Similar conclusions were established by \cite{Mckeon2023resolvent} using resolvent-based estimation with correct second-order flow statistics.
Deep-learning approaches inspired by computer vision have also been applied to reconstruct the near-wall turbulence from planar observations in the outer layer, and the error of the predicted wall stress and pressure were within 10\% of local fluctuations \citep{balasubramanian2023cnn}. 
Building upon these investigations, a fundamental question arises: Given the spatio-temporally resolved velocity data throughout the entire outer layer, what is the highest attainable estimation accuracy of near-wall turbulence?  
In addressing this question, we are particularly interested in predictions that satisfy the Navier-Stokes equations, which was not a condition in the above quoted studies.

Data assimilation techniques, especially when they strongly enforce the governing Navier-Stokes equations, are powerful tools for exploring the fundamental limitation of turbulence reconstruction \citep{Zaki2021da}.
The most intuitive approach to combine observations into equations is continuous data assimilation, which augments the equations with a forcing term that drives the estimated state towards observations \citep{Eyink2013,PCDL2020}.
This approach has been adopted to identify the critical thickness of an unknown wall-attached horizontal layer that can be synchronized, given outer fully resolved velocity data \citep{Wang2022synch}.
If the amount of observations is insufficient for a perfect reconstruction, ensemble-variational methods \citep[EnVar,][]{Liu2008envar} or adjoint-variational approaches \citep[4DVar,][]{Dimet1986_4dvar} are better alternatives to explore the highest possible estimation accuracy.
Both methods aim at minimizing a cost function defined as the difference between observations and their estimation, with the evolution of estimated state being strictly required to satisfy the Navier-Stokes equations.
The minimization of the cost function is distinct in ensemble and adjoint variational methods: The former approximates the gradient of the cost function using an ensemble of forward simulations \citep{Mons2017sensor,Buchta2021envar}; The latter computes the gradient by solving the adjoint equations \citep{zaki2025}.
When applied to reconstructing high-dimensional unknown parameters, such as initial or boundary conditions, the adjoint approach has been demonstrated to be more accurate and efficient than other data-assimilation methods \citep{Mons2016}.
\citet{Wang2021} applied 4DVar to reconstruct turbulent channel flow from  velocity data that were systematically sub-sampled in three-dimensional space and time, throughout the flow domain. 
The data resolutions were comparable to experimental measurements, and they demonstrated that the accuracy of the estimated flow was robust against measurement noise.
In addition, the adjoint field provides the exact sensitivity of observations to the flow state, and thereby sheds light on the causality in turbulence dynamics.
Therefore, we adopt an adjoint-variational approach to construct a Navier-Stokes solution that best reproduces velocity observations in the outer layer while generating all the missing scales in the wall layer.
Such a solution, once obtained, provides a quantitative assessment of the optimal estimation of the near-wall turbulence from outer measurements.
Another interpretation is that the assimilation identifies the missing portion of the near-wall flow that influences the outer observed scales.
We will also explore the influence of Reynolds number and data resolution on the estimation accuracy.  

\begin{figure}[t]
	\centering
	\includegraphics[width=0.6\textwidth]{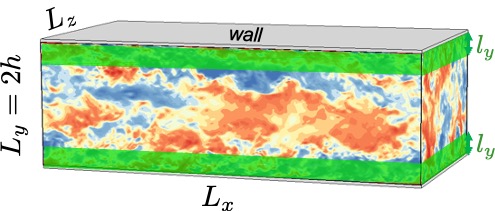}
	\caption{Schematic of turbulent channel flow. The two green regions mark the wall-attached horizontal layers that are unobserved.
	}
	\label{fig:schematic}
\end{figure}

In \S\ref{sec:method}, we introduce the adjoint-variational data assimilation method, and provide details of the computational setup.
The data-assimilation results are presented in \S\ref{sec:results}, all using observations extracted from direct numerical simulations (DNS).
We first consider fully resolved velocity observations in the outer layer and attempt to reconstruct turbulence in the unknown wall layer.
The performance of the adjoint-variational approach is examined in \S\ref{sec:ly50}.
The estimation accuracy is analyzed in Fourier space to identify the scales of near-wall flow structures that can be decoded from outer observations.
In \S\ref{sec:ly90}, we explore a range of thicknesses of the unknown wall layer and of the Reynolds number.
A comparison against linear correlation-based methods is provided, followed by a summary of the highest attainable estimation accuracy of wall quantities using outer observations.
In \S\ref{sec:filtered_data}, we adopt under-resolved velocity data in the outer layer, and discuss the effect of data resolution on the optimal reconstruction of near-wall turbulence.
The conclusions drawn from these tests are summarized in \S\ref{sec:conclusions}.

\section{Adjoint-variational data assimilation}
\label{sec:method}

The flow configuration of interest is a rectangular channel, as shown in figure \ref{fig:schematic}.
The fluid motion is periodic in the streamwise ($x$) and spanwise ($z$) directions, and bounded by two parallel no-slip walls in the vertical direction ($y$).
The reference length and velocity scales are the half height of the channel $h^*$ and the bulk velocity $U_b^*$, also referred to as outer scales, where the superscript $^*$ denotes dimensional quantities.
Another set of normalization scales are the kinematic viscosity $\nu^*$ and the friction velocity $u_{\tau}^* \equiv \sqrt{\langle \tau_{\textrm{w}}^* \rangle / \rho^*}$, where $\langle \tau_{\textrm{w}}^* \rangle$ is the mean shear stress on the walls and $\rho^*$ is the density.
When quantities are normalized by these inner scales, they are marked with the superscript $^+$.
Based on the outer and inner scales, the bulk and friction Reynolds numbers are defined as $Re \equiv U_b^* h^* / \nu^*$ and $Re_{\tau} \equiv u_{\tau}^* h^* / \nu^*$, respectively.

Given velocity observations in the outer layer, $\Omega_{O} \equiv \left\{\boldsymbol{x} \ | \ y \in [l_y, 2-l_y] \right\}$, our objective is to reconstruct all the scales of turbulence in the remaining unobserved wall layers, $\Omega_I \equiv \left\{\boldsymbol{x} \ | \ y \in [0,l_y) \cup (2-l_y,2] \right\}$, which are marked in green in figure \ref{fig:schematic}.
In order to have full control of data resolution and examine the prediction accuracy, we adopt observations generated from direct numerical simulations.
Within the entire computational domain $\Omega = \Omega_O \cup \Omega_I$, we assume the estimated flow state exactly satisfies the Navier-Stokes equations, $\boldsymbol{u}_{n+1} = \mathcal{N}(\boldsymbol{u}_{n})$, where $\boldsymbol{u}_n$ is the velocity field at the $n^{\textrm{th}}$ time step.
As such, the only unknown is the initial condition $\boldsymbol{u}_0$, which can be reconstructed by minimizing a cost function,
\begin{equation}
    \label{eq:cost}
    \mathcal{J}(\boldsymbol{u}_0) = \sum_{n=0}^N \frac 12 \|\boldsymbol{m}_n - \mathcal{M}(\boldsymbol{u}_n) \|^2.
\end{equation}
The right-hand side of (\ref{eq:cost}) quantifies the difference between the instantaneous measurements $\boldsymbol{m}_n$ and their estimation from an initial condition $\boldsymbol{u}_0$.
The notation $\mathcal{M}$ represents the observation operator that generates the measured quantity from the velocity field $\boldsymbol{u}_n$.
In the adjoint-variational approach, the cost function is minimized using a gradient-based optimization algorithm, and the gradient is calculated by solving the adjoint equations.
In the following, we introduce the detailed forward model, adjoint equations, and the optimization algorithm.

\subsection{Forward equations and observations}
\label{sec:forward}

The evolution of the forward velocity vector $\boldsymbol{u}$ and pressure $p$ is governed by the non-dimensional incompressible Navier-Stokes equations,
\begin{subequations}
\label{eq:NS}
    \begin{eqnarray}
        \frac{\partial \boldsymbol{u}}{\partial t} + \left(\boldsymbol{u} \cdot \boldsymbol{\nabla}\right) \boldsymbol{u} &=& -\boldsymbol{\nabla} p + \frac{1}{Re}\nabla^2 \boldsymbol{u}, \\
        \boldsymbol{\nabla} \cdot \boldsymbol{u} &=& 0,  
    \end{eqnarray}
\end{subequations}
where $t$ represents time.
These equations are solved using a second-order fractional-step method with a local volume-flux formulation on a staggered grid \citep{Rosenfeld1991}.
The advection terms are discretized explicitly in time using the second-order Adams-Bashforth scheme, and the diffusion terms are treated implicitly with the Crank-Nicolson scheme.
The pressure Poisson equation is solved using Fourier transforms in the periodic wall-parallel directions, followed by a tri-diagonal inversion in the wall-normal direction.
These numerical schemes have been validated and used extensively for direct numerical simulations (DNS) of wall turbulence \citep{Zaki2013,Jelly2014,Marxen_2019}.

The true, or reference, flow state $\boldsymbol{u}_r$ is statistically stationary turbulence, and is sustained by a constant streamwise pressure gradient, which is always assumed to be known for the purposes of the data assimilation.
We consider two Reynolds numbers, $Re_{\tau} = \{392, 590\}$, and focus on the latter for the majority of our discussion.
A Cartesian grid is adopted with uniform spacing in the streamwise and spanwise directions and hyperbolic tangent stretching in the wall-normal direction.
The domain sizes and grid resolutions are summarized in table \ref{table:Re}.
In each case, the time-step size $\Delta t$ is chosen to guarantee that the Courant–Friedrichs–Lewy number is less than one half.
The flow statistics have been validated against previous research of turbulent channel flow at the same Reynolds numbers \citep{Moser1999}.

Observations of all three components of the velocity are extracted from the reference simulation $\boldsymbol{u}_r$.
In order to minimize the influence of data resolution and focus on exploring the highest estimation accuracy of near-wall turbulence, we start with velocity data that are available at every time step and every grid point within the outer layer $\Omega_O$.
The corresponding observation operator $\mathcal{M}$ is defined as $\mathcal{M}(\boldsymbol{u}(\boldsymbol{x})) = \boldsymbol{u}(\boldsymbol{x})|_{\boldsymbol{x} \in \Omega_O}$
The observation time horizon is assumed to be sufficiently long such that the estimated flow state reaches statistical stationarity, while reproducing the available instantaneous observations.
Three thicknesses of the unknown layers $\Omega_I$ are considered, $l_y^+ = \{50,70,90\}$, which correspond to $l_y = \{0.085,0.12,0.15\}$ at $Re_{\tau} = 590$.
All these thicknesses exceed the threshold $l^+_{y,c} \approx 30$ that guarantees synchronization of near-wall turbulence to outer observations.
According to the law of the wall, the unknown regions include the viscous sublayer, the buffer layer, and the lower part of the logarithmic layer, and as such the region where the majority of the production of turbulent kinetic energy is concentrated.
In other words, the assimilation must reconstruct the engine that is responsible for the generation of the majority of the turbulence kinetic energy from outer observations, which is challenging.
In addition, two of the thicknesses that we consider are beyond $10\%$ of the half-channel height, which is consistent with the typical height of the first off-wall grid point used in wall modelling for large-eddy simulations \citep{Piomelli2008wmles,Larsson2016wmles}.
As such, our observation setup is designed to elucidate the maximum achievable accuracy in predicting wall stresses, given velocity data away from the wall.
Turbulence reconstruction using under-resolved data is attempted in \S\ref{sec:filtered_data}, where we provide more specific information regarding the relevant coarse-graining of the observations.

\begin{table}
	\centering
	\begin{tabular}{c c c c c c c c c c c}
	    \hline
		$Re_{\tau}$ & $Re$ & $L_x/h$ & $L_z/h$ & $N_x$ & $N_y$ & $N_z$ & $\Delta x^+$  & $\Delta y^+_{min}$ & $\Delta y^+_{max}$ & $\Delta z^+$ \\
		\hline
		392 & ~6{,}875 & $2\pi$ & $\pi$ & 257 & 321 & 193 & 9.6 & 0.34 & 5.1 & 6.4\\
		\rowcolor{blue!10} 590 & 10{,}935 & $2\pi$ & $\pi$ & 385 & 385 & 385 & 9.6 & 0.44 & 6.5 & 4.8 \\
		\hline
	\end{tabular}
	\caption{Computational domains and grid sizes for simulations at different Reynolds numbers.}
	\label{table:Re}
\end{table}

\subsection{Adjoint-variational data assimilation}
\label{sec:adjoint}

Under the Navier-Stokes constraint (\ref{eq:NS}), any perturbation to the initial condition $\boldsymbol{u}_0$ will affect all the subsequent flow fields and the estimated observations $\mathcal{M}(\boldsymbol{u}_n)$.
As such, the corresponding variation in the value of the cost function (\ref{eq:cost}) is non-trivial to calculate.
In the limit of infinitesimal perturbation, we can derive the gradient of the cost function with respect to the initial condition, $\mathscr{D} \mathcal{J} / \mathscr{D} \boldsymbol{u}_0$, by introducing the Lagrangian,
\begin{equation}
    \label{eq:Lagrangian}
    \mathcal{L} = \mathcal{J} - \left(\boldsymbol{u}^{\dag},  \frac{\partial \boldsymbol{u}}{\partial t} + \boldsymbol{u} \cdot \nabla \boldsymbol{u} + \nabla p - \frac{1}{Re}\nabla^2 \boldsymbol{u}\right) - \left(p^{\dag}, \nabla \cdot \boldsymbol{u} \right),
\end{equation}
where the Lagrangian multipliers $(\boldsymbol{u}^{\dag},p^{\dag})$ are also termed as adjoint velocity and pressure, respectively.
The brackets in equation (\ref{eq:Lagrangian}) denote the inner product between two vector fields, defined as an integration over the entire domain $\Omega$ and the assimilation time horizon $t\in[t_s,t_e]$,
\begin{equation}
    \label{eq:inner_product}
    \left(\boldsymbol{f}, \boldsymbol{g} \right) = \int_{t_s}^{t_e} \int_{\Omega} \boldsymbol{f} ^{\top} \boldsymbol{g} \ \mathrm{d}^3 \boldsymbol{x} \mathrm{d}t.
\end{equation}
Due to the inner-product terms in the definition (\ref{eq:Lagrangian}), the Lagrangian is a function of both the forward variables $\boldsymbol{q} = [\boldsymbol{u},p]^{\top}$ and their adjoints $\boldsymbol{q}^{\dag} = [\boldsymbol{u}^{\dag},p^{\dag}]^{\top}$.
The first-order optimality condition requires setting the functional derivatives to zero, $\mathcal{D} \mathcal{L} / \mathcal{D} \boldsymbol{q}^{\dag} = 0$ and $\mathcal{D} \mathcal{L} / \mathcal{D} \boldsymbol{q} = 0$.
The former recovers the forward Navier-Stokes equations (\ref{eq:NS}), while the latter yields the adjoint equations,
\begin{subequations}
    \label{eq:Adjoint}
    \begin{eqnarray}
    \nabla \cdot \boldsymbol{u}^{\dag} &=&  -\frac{\mathcal{D} \mathcal{J}}{\mathcal{D} p}, \\
    \label{eq:Adjoint_mom}
    \frac{\partial \boldsymbol{u}^{\dag}}{\partial t^{\dag}} - \boldsymbol{u} \cdot \nabla \boldsymbol{u}^{\dag} + (\nabla \boldsymbol{u}) \cdot \boldsymbol{u}^{\dag} &=& \nabla p^{\dag} + \frac{1}{Re} \nabla^2 \boldsymbol{u}^{\dag} + \frac{\mathcal{D} \mathcal{J}}{\mathcal{D} \boldsymbol{u}}.
    \end{eqnarray}
\end{subequations}
The functional derivatives $\mathcal{D} \mathcal{J} / \mathcal{D} p$ and $\mathcal{D} \mathcal{J} / \mathcal{D}\boldsymbol{u}$ are derived analytically by assuming that the forward fields at different times are independent.
Since we only consider velocity observations, the functional derivative with respect to the pressure is always zero, $\mathcal{D} \mathcal{J} / \mathcal{D} p = 0$.
The derivative with respect to the velocity $\mathcal{D} \mathcal{J} / \mathcal{D} \boldsymbol{u}$ is notably different from the gradient of the cost function with respect to the initial condition, $\mathscr{D}\mathcal{J} / \mathscr{D} \boldsymbol{u}_0$, which quantifies the sensitivity of the cost function to the initial state when the governing equations are rigorously satisfied.
The notation $t^{\dag} \equiv t_e - t$ in the adjoint momentum equation (\ref{eq:Adjoint_mom}) is the reverse time, which indicates that the adjoint equations are solved backwards in time.
The adjoint advection terms involve the forward velocity field $\boldsymbol{u}$.
Therefore, the time history of the forward velocity must be stored and subsequently retrieved during the adjoint evolution.
At the end of the adjoint marching, the gradient of the cost function is provided by the adjoint field at the initial time $t=t_s$ ($t^{\dag}=t_e - t_s$),
\begin{equation}
    \label{eq:grad}
    \frac{\mathscr{D} \mathcal{J}}{\mathscr{D} \boldsymbol{u}_0} = \boldsymbol{u}^{\dag}_0 \equiv \boldsymbol{u}^{\dag}|_{t = t_s}.
\end{equation}
This gradient is then be used in a gradient-based optimization algorithm to search for the next estimate of $\boldsymbol{u}_0$ that reduces the value of the cost function.
Here we adopt the limited-memory Broyden-Fletcher-Goldfarb-Shanno (L-BFGS) method due to its efficiency and robustness in solving high-dimensional nonlinear optimization problems \citep{LBFGS}.

The above equations (\ref{eq:NS}-\ref{eq:grad}) were derived and expressed in continuous form to facilitate a comparison between the forward and adjoint systems.
Given the discretization of the forward equations in \S\ref{sec:forward}, the choice of the numerical scheme for the discretization of the adjoint equations is non-trivial, and significantly affects the accuracy of the computed gradient of the cost function.
Here the discrete adjoint approach is adopted, where we first formulate the Lagrangian (\ref{eq:Lagrangian}) using the discrete forward equations and then derive the discrete adjoint equations using the first-order optimality conditions \citep{Wang2019}.
This approach guarantees that the discrete adjoint system provides the exact gradient of the cost function.
The computational cost of an adjoint simulation is approximately equal to a forward one. Since the adjoint equations feature the instantaneous forward velocity fields (the second term in equation \ref{eq:Adjoint_mom}), the forward velocity is stored during the Navier-Stokes solution.  A common approach to reduce the storage is checkpointing, where the forward fields are stored at a few checkpoints and the forward simulation is repeated between checkpoints to generate the required velocity fields during adjoint marching \citep{Eggl2018mixing}.

\begin{figure}[t]
	\centering
	\includegraphics[width=0.9\textwidth]{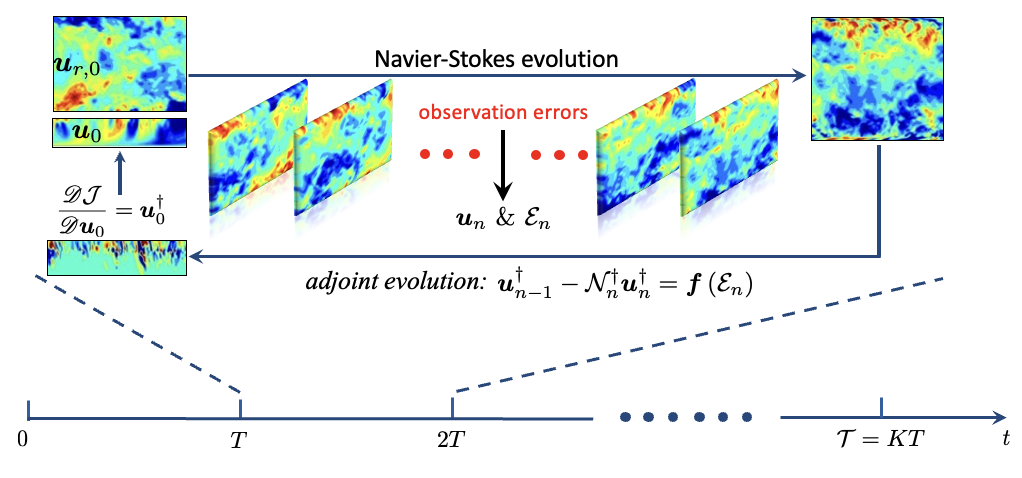}
	\caption{
    Schematic representation of the adjoint-variational data assimilation algorithm with a sliding-window strategy. The observation time horizon $\mathcal{T}$ is divided into $K$ windows, and adjoint data assimilation is performed consecutively in each window. Within each assimilation window, the first estimate of the initial condition is constructed and then advanced forward using the Navier-Stokes equations.  The forward velocity $\boldsymbol{u}_n$ and the deviation from observations $\mathcal{E}_n$ are recorded. The adjoint system is evolved backward in time, driven by observation errors. The initial adjoint field $\boldsymbol{u}_0^{\dag}$ provides the gradient of the cost function, which is used to update $\boldsymbol{u}_0$ within the wall layer. The forward-adjoint loop is repeated until convergence.
	}
	\label{fig:4dvar_schematic}
\end{figure}

Combining the forward equations, adjoint equations, and the gradient-based optimization method, we obtain the complete procedure of adjoint-variational data assimilation (Algorithm \ref{alg:adjoint}), which is also summarized schematically in figure \ref{fig:4dvar_schematic}.
Starting from the first estimate of the initial flow state $\boldsymbol{u}_0$, we solve the forward equations from $t=t_s$ to $t=t_e$, store the instantaneous forward velocity fields, and evaluate the differences between the estimated and available observations.
After the forward simulation is completed, the adjoint equations are advanced backward in time from $t=t_e$ (or $t^{\dag} = 0$) to $t=t_s$ (or $t^{\dag} = t_e - t_s$).
At the end of the adjoint evolution, the adjoint field $\boldsymbol{u}_0^{\dag}$ provides the gradient of the cost function, which is adopted in the L-BFGS algorithm to update our estimate of the initial condition and minimize $\mathcal{J}$.
This forward-adjoint loop is then repeated, either for a prescribed number of iterations or until a specified convergence threshold is reached.
In all the examined data-assimilation cases, the optimization algorithm is performed for 100 L-BFGS iterations, or forward-adjoint loops, to make comparisons at a fixed computational cost.

\begin{algorithm}
	\SetAlgoLined
    \textbf{Initialization}: Construct a first estimate of the initial condition $\boldsymbol{u}_{0}$ at $t = t_s$\;
	\textbf{Step 1}: Forward model\;
	\Indp
	\textbullet~Advance the estimated initial condition $\boldsymbol{u}_{0}$ using the forward equations (\ref{eq:NS}), from $t=t_s$ to $t=t_e$, and store the time-dependent forward velocity fields\;
	\textbullet~Evaluate the cost function (\ref{eq:cost})\;
	\Indm
	\textbf{Step 2}: Adjoint model\;
	\Indp
	\textbullet~March the adjoint equations (\ref{eq:Adjoint}) from $t^{\dag} = 0$ (or $t=t_e$) to $t^{\dag}=t_e - t_s$ (or $t=t_s$)\;
	\textbullet~At the initial time ($t = t_s$), obtain the gradient of the cost function (\ref{eq:grad})\;
	\Indm
	\textbf{Step 3}: Update the estimated initial state\;
	\Indp
	\textbullet~Feed the gradient into the L-BFGS algorithm, and search for an improved estimate of the initial state that reduces the value of the cost function\; 
	\textbullet~Repeat steps 1-3 until convergence or for a prescribed number of iterations. 
	\caption{Adjoint-variational data assimilation.}
	\label{alg:adjoint}
\end{algorithm}

When adjoint-variational data assimilation is applied to turbulent flows, the duration of the assimilation window must be constrained by the Lyapunov timescale \citep{Li2020,Chandramouli2020}.
Specifically, in addition to the chaotic nature of the forward Navier-Stokes dynamics, the adjoint system is also chaotic and amplifies exponentially in backward time at the same Lyapunov exponent \citep{Wang2022Hessian}. 
If the assimilation window is longer than the Lyapunov timescale, machine errors in the forward evolution will amplify appreciably and similarly errors in the representation of the measurements will amplify in the adjoint solution.  In addition to these sources of errors, the exponential amplification of the adjoint field itself will lead to a large gradient which imposes a severe restriction on the step size in gradient-based optimization.
To overcome these difficulties, we adopt the sliding assimilation window, or cycling scheme \citep{Fisher2012}, which is shown in the schematic figure \ref{fig:4dvar_schematic}.
The observation time horizon $[0,\mathcal{T}]$, assumed sufficiently long such that the estimated flow state reaches statistically stationary, is divided into $K$ shorter windows with duration $T = \mathcal{T}/K$. Adjoint-variational data assimilation (Algorithm \ref{alg:adjoint}) is applied within each window consecutively.
Here we choose $T$ to be slightly shorter than the Lyapunov timescale, $T^+ = 0.8 \tau_{\sigma} \approx \{35,31\}$ at $Re_{\tau} = \{392,590\}$ \citep{Nikitin2018}.
The effect of the window size $T$ on the accuracy of the estimated state is examined in Appendix \ref{sec:window}.  We show that shorter windows lead to lower estimation accuracy. Additionally, since 4DVar is used to estimate the velocity field at the start of each time window, the flow state has an adjustment at the boundaries between adjacent windows. Shorter time windows lead to more frequent adjustments.
Starting from the first window, $[t_s,t_e] = [0,T]$, we perform 100 forward-adjoint loops to reconstruct the initial condition at $t=0$.
The estimated initial condition is then marched until $t=T$ to obtain the final state $\boldsymbol{u}(\boldsymbol{x},T)$, which is then adopted as the first guess for the next assimilation window $[t_s,t_e] = [T,2T]$.
The adjoint-variational approach is repeated in $[T,2T]$ and so on until the end of the observation horizon.

The above discussion has focused on reconstructing the initial state within the entire computational domain.
When fully resolved observations of the true state $\boldsymbol{u}_r$ are available in the outer layer, we can set the estimated initial state in that layer to be identical to the observations, and only reconstruct the initial state within the inner layer.
Specifically, at $t=0$, we enforce $\boldsymbol{u}(\boldsymbol{x},0) = \boldsymbol{u}_r(\boldsymbol{x},0)$, for all $\boldsymbol{x} \in \Omega_O$, and the first guess of the flow state in $\Omega_I$ is constructed using linear stochastic estimation (LSE, see Appendix \ref{sec:LSE} for details) from outer observations.
The forward and adjoint equations are still solved within the full domain.
At the end of the adjoint simulation, the gradient with respect to the initial state within the inner layer only is utilized in the optimization algorithm.
As data assimilation is completed in the first window $[0,T]$, we replace the estimated state at $t=T$ in the outer layer by the true observations, and only keep the inner-layer estimation as the first guess for the next window.
The same strategy is repeated for subsequent assimilation windows.

\section{Results}
\label{sec:results}

\subsection{Performance of the algorithm: multiple windows and spectral analysis}
\label{sec:ly50}

\begin{figure}[t]
	\centering
	\includegraphics[width=\textwidth]{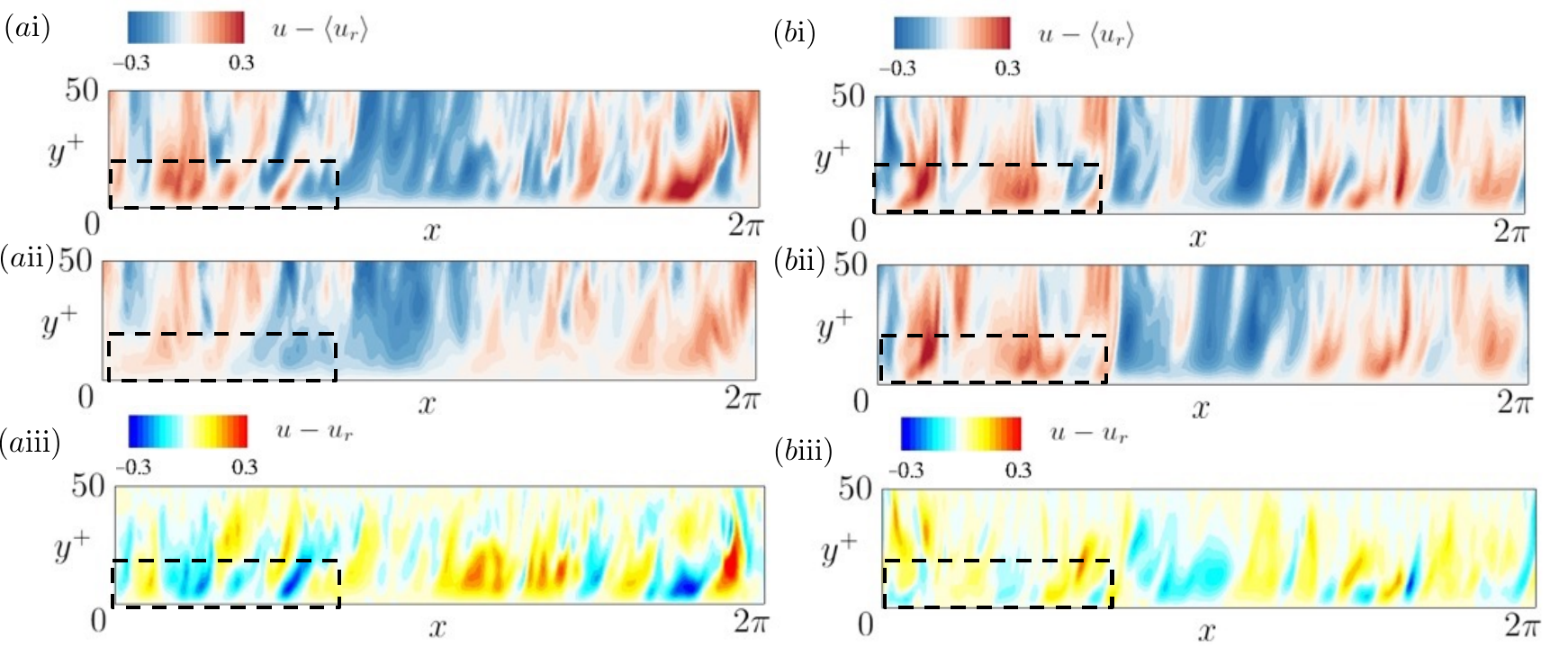}
	\caption{
    Instantaneous streamwise-velocity fluctuations at $z = 0.49L_z$, calculated relative to the true mean.  (i) Adjoint-variational reconstruction; (ii) true fields; estimation errors. Panels ($a$) and ($b$) correspond to the initial ($t=0$) and final ($t=T$) times of the first window. The black boxes highlight the region where turbulence fluctuation is underestimated at $t=0$ but better reconstructed at $t=T$.
	}
	\label{fig:ly50_u_xy_t0_tT}
\end{figure}

In this section, we focus on results from channel flow at $Re_\tau = 590$.  
We start by considering fully resolved velocity observations outside $l_y^+ = 50$.
The unknown wall-attached layer includes the peak turbulent kinetic energy (TKE) and the majority of the contributions to the TKE budget, such as production, dissipation, and turbulent transport.
In terms of vorticity dynamics, no information about the flux of vorticity at the wall is available in the observations nor of the intense near-wall stretching and ejection events that lead to the extrema in the wall stresses \citep{Eyink2020_channel,Wang2022SCauchy}.
In addition, beyond $y^+ = 50$, the transport of turbulent enstrophy almost vanishes \citep{Gorski_Wallace_1994}.
Since an important dynamical region of the channel is absent from the observations, an accurate reconstruction of the turbulence is expected to be challenging.

A qualitative perspective on the adjoint-variational estimated state is provided in figure \ref{fig:ly50_u_xy_t0_tT}(i), for the first assimilation window $t \in [0, T]$.
At the initial time $t=0$, the estimated streamwise fluctuations (panel $a$i) near the first observation plane at $y^+=50$ closely match the true state (panel $a$ii).
At locations more distant away from the first observation plane, and closer to the wall, the estimation quality deteriorates, as highlighted within the black dashed boxes in panels ($ai$) and ($aii$).
The magnitude of the fluctuations is underestimated, due to the lack of sensitivity of observations to the initial state in the near-wall layer.
The difference between the estimated and true field is quantified in panel ($aiii$), which clearly shows the increase in errors away from the first measurement point.
As the estimated state evolves forward using the Navier-Stokes equations, there is a noticeable improvement in the reconstructed turbulent kinetic energy and small-scale motions, which is highlighted within the dashed boxes in panels $bi$ and $bii$.
The estimation error is also reduced from $t = 0$ to $t = T$, as shown in panel $biii$.
The duration of a single observation window, $T^+ = 31$, may be insufficient for the estimated near-wall flow to reach a statistically stationary state.  In addition, this time horizon may also be insufficient for all the scales near the wall to affect the outer flow, and hence the observations.  For these reasons, data assimilation across multiple windows has the potential to further improve the estimation accuracy.

\begin{figure}[t]
    \centering
    \includegraphics[width=\textwidth]{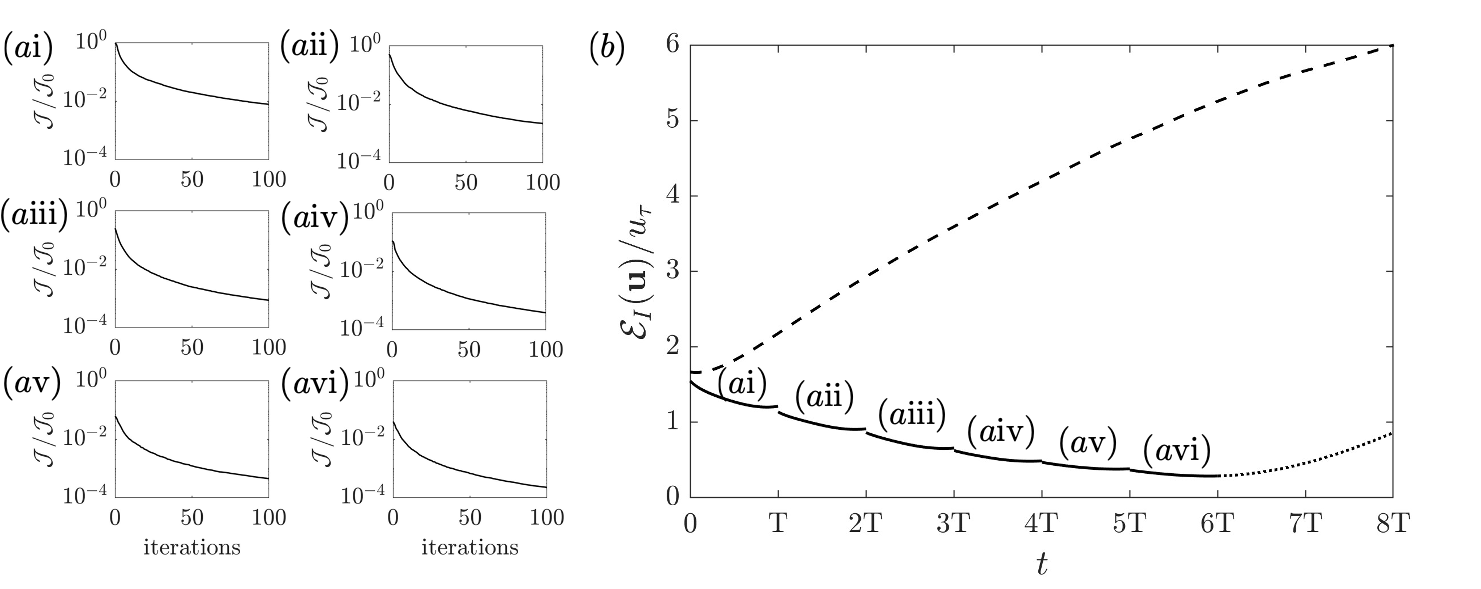}
    \caption{
    ($a$) Convergence history of the cost function when adjoint-variational approach is applied to six consecutive observation windows separately ($T = 0.96$, $T^+ = 31$). In all six panels, the cost function is normalized by the initial value of the first window, $\mathcal{J}_0$.
    ($b$) Volume-averaged error in the  velocity versus time, for the Navier-Stokes evolution of (dashed) the first guess, (solid) the adjoint-variational estimated state, and (dotted) the predicted flow starting from the final observation time $t = 6T$.
    }
    \label{fig:ly50_errv_cost}
\end{figure}

The adjoint-variational data assimilation is repeated over six consecutive windows, which are sufficient to ensures that the estimated near-wall turbulence reaches a statistically stationary state.
The convergence histories of the cost functions in these windows are shown in figure \ref{fig:ly50_errv_cost}$a$i-$a$vi.
All the cost functions are normalized by the value from the first guess in the first window, $\mathcal{J}_0$.
Within each assimilation window, after 100 forward-adjoint loops, the cost function always drops by more than two orders of magnitude. At the start of each subsequent window, however, the cost increases relative to the final iteration of the preceding one, which is expected because the new observations were not at all taken into account in the preceding assimilation.  Nonetheless, the overall trend is a progressive decrease in the cost function, which at the end of the sixth window reaches $10^{-4}\mathcal{J}_0$.
Recall that the cost function is defined as the kinetic energy norm of the errors in reproducing the observations.  As such, the cost reduction by the 4DVar algorithm translates to a reduction in the errors of reproducing the observed velocities to 1\% of the initial values. 

Beyond reproducing observations, it is crucial to evaluate the performance of the adjoint-variational approach in the near-wall layer where observations are not available.
The estimation accuracy is quantified using the root-mean-square (r.m.s.) error between the estimated and true states,
\begin{equation}
    \label{eq:err_V}
    \mathcal{E}_I(\boldsymbol{u}) = \langle \left\| \boldsymbol{u} - \boldsymbol{u}_r\right\|^2 \rangle_{\Omega_I}^{1/2},
\end{equation}
where $\langle \cdot \rangle$ denotes averaging, and the subscript represents the averaging domain.
The temporal evolution of the estimation error is plotted in figure \ref{fig:ly50_errv_cost}$b$.  The dashed line reports the outcome of advancing the initial guess, which is obtained by linear stochastic estimation at $t=0$, using the Navier-Stokes equations; The solid line is the outcome of performing the sliding-window adjoint reconstruction within the six time intervals.
Starting with the LSE result, the estimation error monotonically increases, which is a manifestation of the chaotic nature of turbulence. In contrast, the error of the adjoint-estimated state progressively decreases, demonstrating the convergence to the true turbulent state.
If the estimated state at $t = 6T$ is marched forward using the Navier-Stokes equations, the error starts to grow due to turbulence chaos (dotted line in figure \ref{fig:ly50_errv_cost}$b$). Despite the divergence of the predicted flow from the true state, the accuracy of the forecast remains comparable to that within the assimilation horizon for a short time.

\begin{figure}[t]
	\centering
	\includegraphics[width=\textwidth]{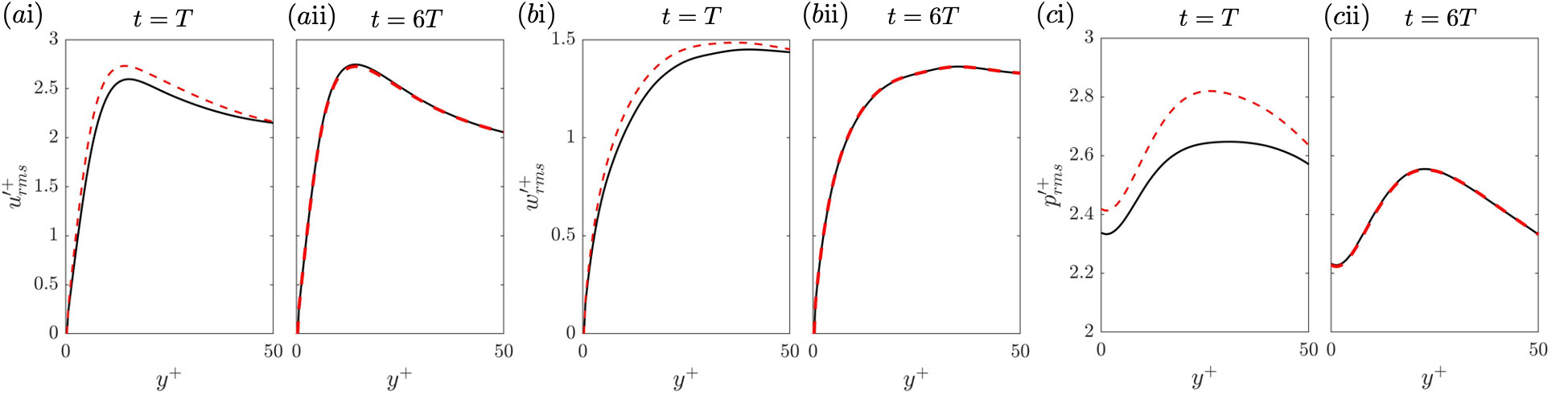}
	\caption{Comparison of the instantaneous horizontally-averaged flow statistics between (black solid) adjoint-variational estimation and (red dashed) true field at (i) $t = T$ and (ii) $t = 6T$. ($a$,$b$,$c$) R.m.s. fluctuations of streamwise velocity, spanwise velocity and pressure, evaluated by subtracting their true mean.
	}
	\label{fig:ly50_rms}
\end{figure}

\begin{figure}[t]
	\centering
	\includegraphics[width=\textwidth]{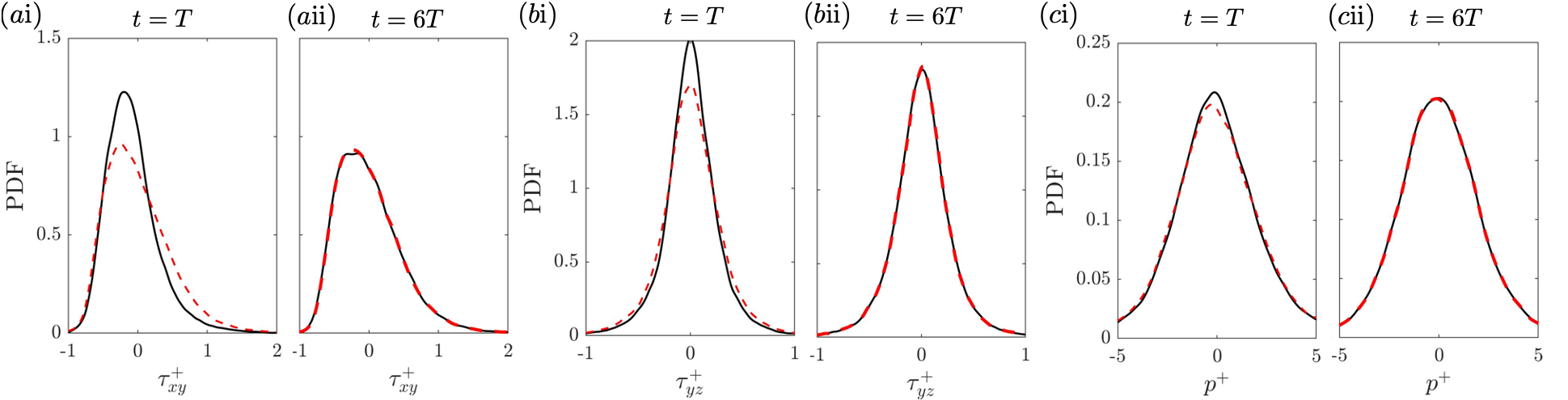}
	\caption{Comparison of the probability density function of wall signals between (black solid) adjoint-variational estimation and (red dashed) true field at (i) $t = T$ and (ii) $t = 6T$. ($a$,$b$,$c$) Fluctuations of wall shear stress $\tau_{xy}^{\prime}$, $\tau_{yz}^{\prime}$ and pressure $p^{\prime}$, calculated by subtracting the true mean and normalized by the true mean shear stress at the wall $\langle \tau_{\textrm{w}} \rangle$. 
	}
	\label{fig:ly50_PDF}
\end{figure}

In order to determine whether the estimated state reaches statistical stationarity in the wall layer, we examine the wall-normal profiles of the root-mean squared (r.m.s.) fluctuations in figure \ref{fig:ly50_rms}, and the probability density function (PDF) of the wall shear stresses and pressure in figure \ref{fig:ly50_PDF}.
At $t=T$, the reconstructed fluctuations of velocity and pressure are all weaker than the true state (figure \ref{fig:ly50_rms}(i)), which aligns with the qualitative comparison in figure \ref{fig:ly50_u_xy_t0_tT}.
Additionally, the PDFs of the estimated wall signals exhibit narrower tails than the true profiles, which indicates a reduced probability of the extreme events at $t=T$.
At the end of the sixth window, $t=6T$, the reconstructed fluctuations and PDFs closely match the true statistics.
Combined with the diminishing estimation error in figure \ref{fig:ly50_errv_cost}, these results validate that the adjoint-variational algorithm successfully converges to a statistically stationary turbulent state that shadows the true flow state in the outer layer and reproduces its observations.

\begin{figure}[t]
	\centering
	\includegraphics[width=\textwidth]{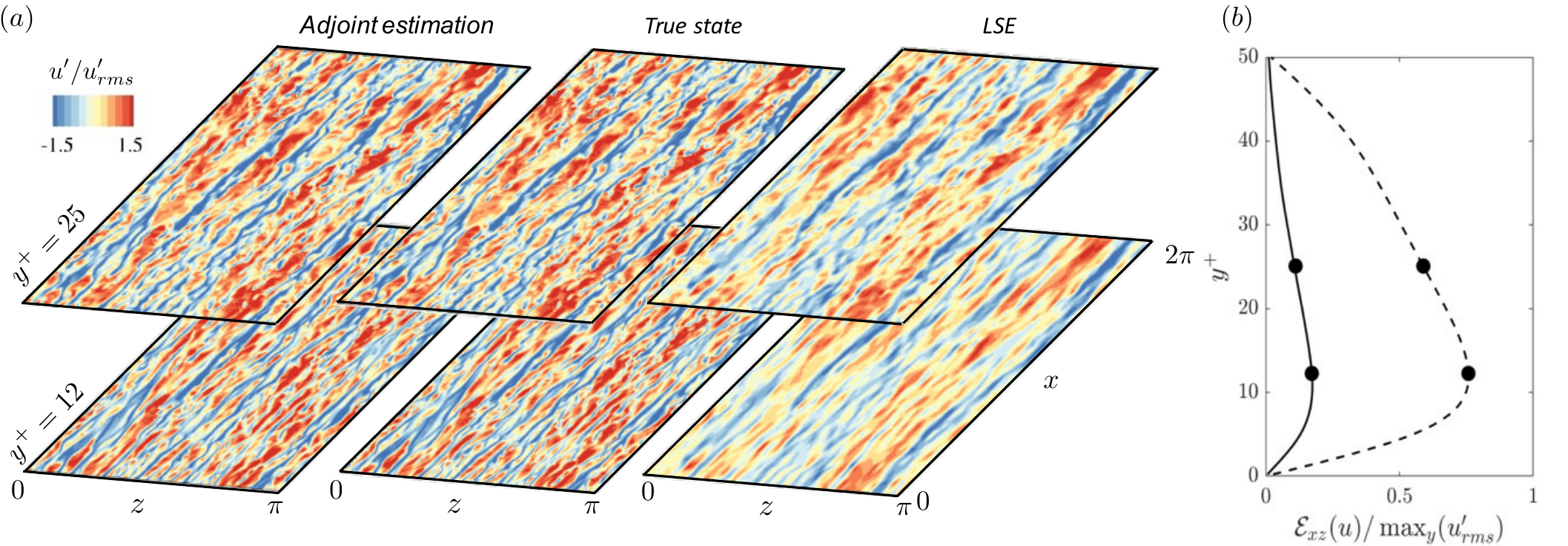}
	\caption{($a$) Streamwise velocity fluctuations in the adjoint-variational estimated field, true state, and linear stochastic estimation, at $t = 6T$. The visualized fields are normalized by the true r.m.s. fluctuations at the corresponding wall-normal location.
    A version of this figure featured in the recent review by \citet[][CC BY 4.0]{zaki2025}.
    ($b$) Horizontally-averaged estimation error of streamwise velocity, reconstructed using (dashed) LSE and (solid) adjoint approach, normalized by the maximum value of r.m.s. streamwise fluctuations.
	}
	\label{fig:ly50_ucontour_erru}
\end{figure}

The adjoint-variational reconstruction of streamwise fluctuation at $t=6T$ is visualized in \ref{fig:ly50_ucontour_erru}$a$, with a comparison against the true field and a linear stochastic estimation.
Here the LSE is performed using contemporaneous outer observations, at $t=6T$.
Inclusion of earlier observations into LSE could improve the estimation accuracy, but previous research has explored this strategy and concluded such enhancements are minimal \citep{Jimenez2019LSE}.
The two visualized horizontal planes are at the midpoint of the unknown layer ($y^+ = 25$) and the location of peak turbulent kinetic energy ($y^+ = 12$).  At both planes, the adjoint estimation reproduces all the scales of the near-wall turbulence, while LSE only captures the large-scale motions.
A quantitative assessment is provided in panel $b$, which presents the wall-normal profiles of the horizontally-averaged estimation errors.
The largest error using LSE is 75\% of the maximum r.m.s.~fluctuations, which implies that only 25\% of the near-wall turbulent kinetic energy can be estimated from outer motions when linear methods are adopted.
The error of the adjoint-estimated field is only 17\% of the maximum r.m.s.~fluctuations, which is 77\% lower than LSE and demonstrates the benefit of turbulence reconstruction using the full nonlinear Navier-Stokes equations.

\begin{figure}[t]
	\centering
	\includegraphics[width=\textwidth]{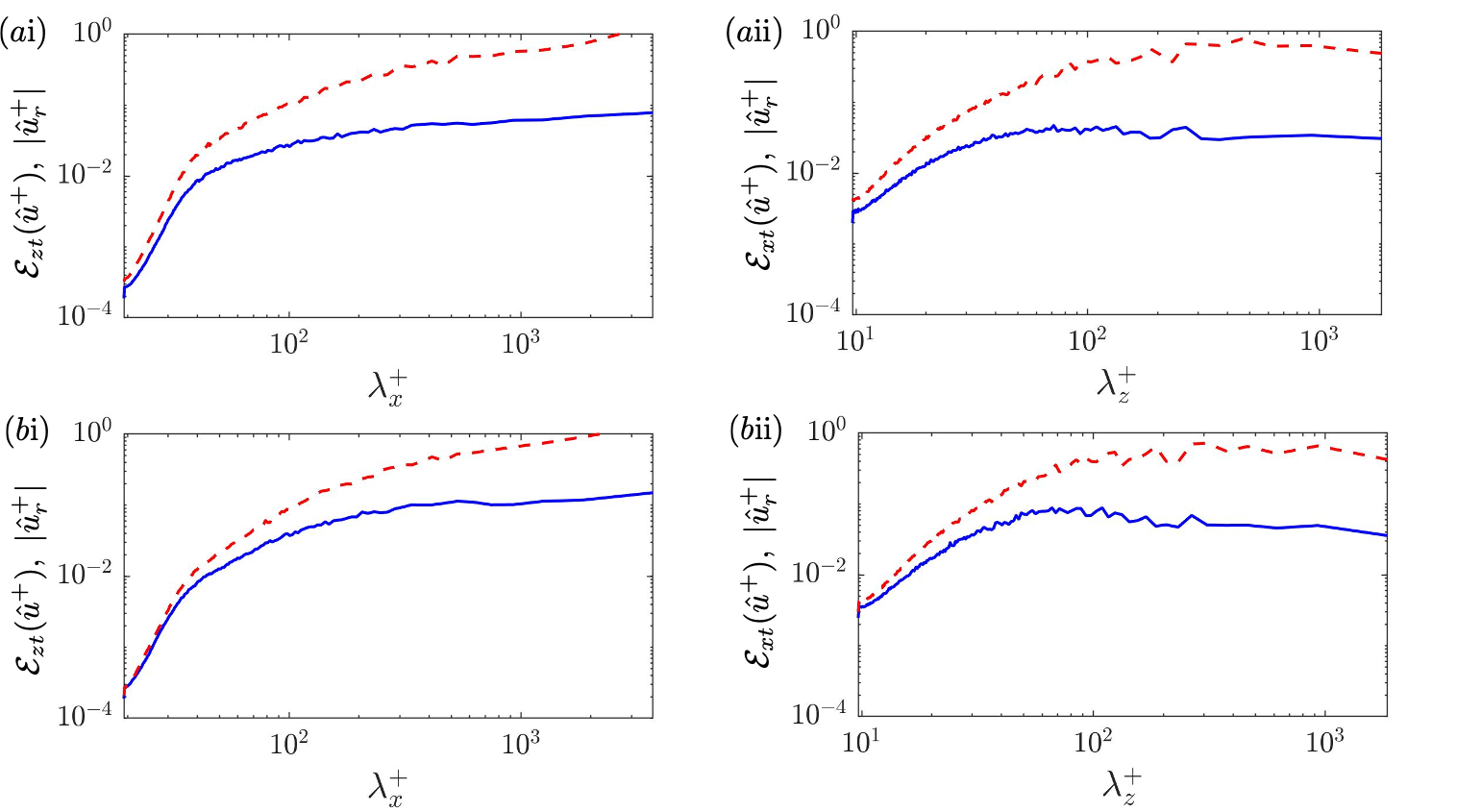}
	\caption{One-dimensional Fourier spectrum of (blue solid) estimation error and (red dashed) the true streamwise velocity, averaged within $t \in [5T, 6T]$, evaluated at ($a$,$b$) $y^+ = \{25, 12\}$.
    (i) Streamwise spectra averaged in the span; 
    (ii) spanwise spectra averaged along the streamwise direction.
	}
	\label{fig:ly50_spectrum_1D}
\end{figure}

To further evaluate the estimation accuracy at different scales, we compute the streamwise spectrum of the adjoint-variational estimation error,
\begin{equation}
    \label{eq:err_Fourier}
    \mathcal{E}_{zt}(\hat{u}) = \langle | \hat{u} - \hat{u}_r |^2\rangle^{1/2}_{zt}, 
\end{equation}
where $\hat{u}$ is the Fourier transform of $u$ in the horizontal directions.  In the above expression, the error is averaged along the spanwise direction and within the last assimilation window $t\in [5T,6T]$.
Similarly, the spanwise spectrum $\mathcal{E}_{xt}(\hat{u})$ is evaluated by averaging the error along the streamwise direction.
The resulting spectra at $y^+ = \{25, 12\}$ are plotted as blue solid lines in figures \ref{fig:ly50_spectrum_1D}($a$ and $b$), respectively.
Although the estimation errors  are larger for longer wavelengths, this trend must be viewed relative to the high energy content in these scales, as shown by red dashed lines.
Theoretically, if the estimated state is fully developed and independent of the truth, the estimation error would equal $\sqrt{2}$ times the true spectra.
This argument follows from decomposing the mean-square error,
\begin{equation}
    \label{eq:error_decompose}
    \mathcal{E}_{zt}^2(\hat{u}) = \langle |\hat{u}-  \hat{u}_r|^2 \rangle_{zt} = \langle |\hat{u}|^2 \rangle_{zt} + \langle |\hat{u}_r|^2 \rangle_{zt} - 2  \langle |\hat{u}^* \hat{u}_r| \rangle_{zt} = 2\langle |\hat{u}_r|^2 \rangle_{zt},
\end{equation}
where the last equality used the assumption of a fully developed estimated state (as such $\langle |\hat{u}|^2 \rangle_{zt} = \langle |\hat{u}_r|^2 \rangle_{zt}$) which is independent of the true flow ($\langle \hat{u}^* \hat{u}_r \rangle_{zt} = 0$). 
Conversely, if the estimated state is correlated with the truth, the error is lower than $\sqrt{2}$ times the true spectra.
Our results in figure \ref{fig:ly50_spectrum_1D} show that the adjoint estimation error is an order of magnitude smaller than the true spectra at the energy-containing large scales, implying a strong positive correlation between the estimated and true large-scale motions.
The error reduction is less significant for smaller scales that represent the dissipative eddies, and at locations farther from observations.
Nevertheless, the estimation error remains below the true spectrum across all the scales, which indicates that even the smallest Kolmogorov eddies in the reconstructed wall layer are correlated with the true flow.
These results demonstrate the capability of our approach to reconstruct all the missing scales of near-wall turbulence, while reproducing velocity observations in the outer layer.

\begin{figure}[t]
	\centering
	\includegraphics[width=\textwidth]{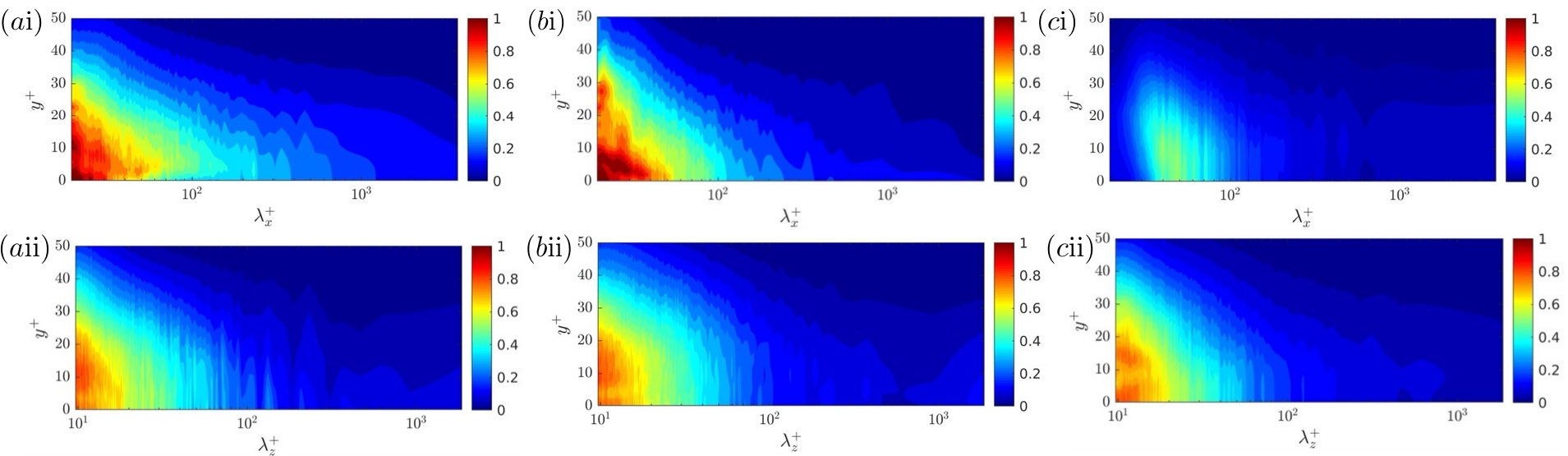}
	\caption{Wall-normal distribution of the one-dimensional Fourier spectrum of estimation error, normalized by the spectrum of the true field. ($a$) Streamwise velocity, ($b$) spanwise velocity and ($c$) pressure. (i,ii) Streamwise and spanwise spectra. The thickness of the unknown layer is $l_y^+ = 50$.
    Figures ($ai$) and ($aii$) featured in the recent review by \citet[][CC BY 4.0]{zaki2025}.
	}
	\label{fig:ly50_spectrum_uwp}
\end{figure}

\begin{figure}[t]
	\centering
	\includegraphics[width=\textwidth]{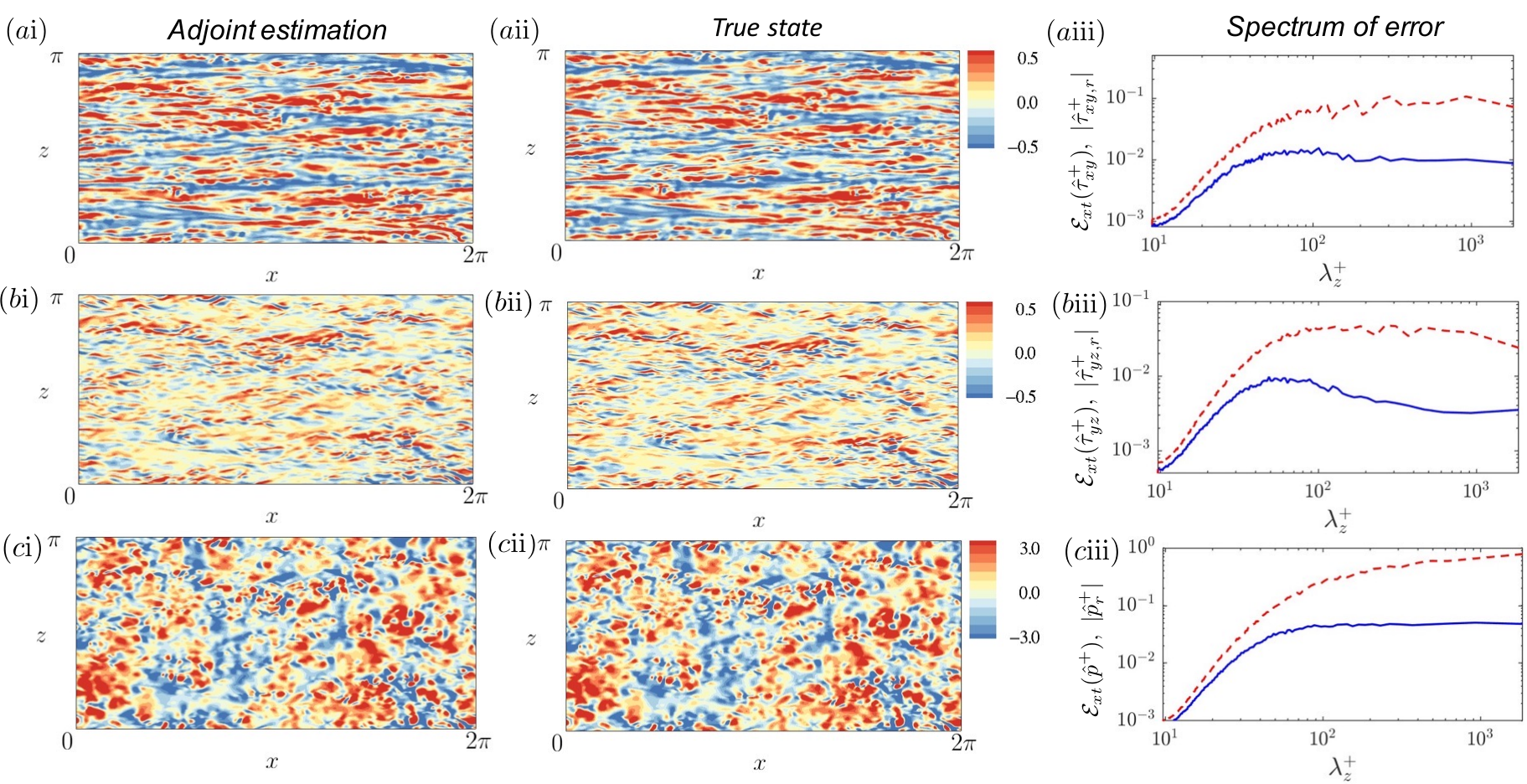}
	\caption{(i) Adjoint-variational estimation of fluctuations of wall shear stress and pressure at $t = 6T$ when $l_y^+ = 50$. (ii) The true fields. (iii) (Blue) spanwise spectra of the estimation error of wall quantities averaged in $t \in [5T,6T]$, compared with (red dashed) the spectra of the true field. ($a$,$b$,$c$) two components of the wall shear stress and pressure, $\{\tau_{xy}^+, \tau_{yz}^+,p^+\}$.
	}
	\label{fig:ly50_vis_wall_spectrum}
\end{figure}

The wall-normal dependence of the error spectra is reported in further detail in figure \ref{fig:ly50_spectrum_uwp}, for the streamwise velocity, spanwise velocity, and pressure.
To ensure a consistent comparison across different flow quantities, the spectra of error are normalized by the true spectra of the corresponding variable.
A key observation is that the large-scale motions, such as quasi-streamwise rolls, are accurately estimated throughout the entire unknown layer, while the accuracy of small scales deteriorates with the distance from the observed regions.
These results are qualitatively consistent with the characteristics of the coherent structures in wall turbulence \citep{Jimenez2018coherent}: longer eddies are also deeper in the wall-normal direction, and the depth saturates until the eddies become attached to the wall.
However, different from the forward perspective that focuses on the evolution of coherent structures, the adjoint approach reveals the turbulent dynamics from a dual, or inverse, perspective.
To construct a Navier-Stokes solution that reproduces outer observations, the large scales across the entire wall layer and the small scales in the vicinity of outer observations must be accurately reconstructed.
The remaining flow structures, especially the small scales far from the outer layer, are only partially observable from the outer flow.

We conclude this discussion by visualizing the shear stresses and pressure at the wall (figure \ref{fig:ly50_vis_wall_spectrum}), which is the most distant location from observations in the observed outer layer.
For all three components, the adjoint reconstruction (panel i) is almost identical to the true state (panel ii), despite slight mismatch in the small scales, as quantified by the spectra of the estimation error (panel iii).
Recall that the thickness of the unknown layer, $l_y^+ = 50$, exceeds the threshold that guarantees a perfect synchronization.
Although the estimation error cannot converge to machine zero, the reconstructed near-wall turbulence is sufficiently accurate to evaluate any flow quantity of interest, such as pressure gradients or vortical structures.

\subsection{The influence of the unobserved layer thickness \\ and of the Reynolds number}
\label{sec:ly90}

When the thickness of the unobserved layer is increased and the observations are more separated from the wall, we can anticipate that an accurate reconstruction of the near-wall turbulence becomes more difficult.
From the perspective of the adjoint-variational optimization algorithm, fewer observations imply that a greater number of Navier-Stokes solutions can replicate the available data with an acceptable level of accuracy.
Accordingly, the landscape of the cost function 
may be expected to feature numerous local minima that can forestall the optimization algorithm.
Moreover, the first guess of the initial condition, which is constructed using linear stochastic estimation, deviates more substantially from the true state, which exacerbates the difficulty of convergence to the global minimum of the cost function.
From the perspective of turbulence dynamics, an elevated $l_y$ results in fewer wall-attached eddies being captured in the observations \citep{Marusic2019review}, leaving more near-wall flow structures unobservable from outer measurements. 
In order to quantitatively determine the impact of $l_y$ on the estimation accuracy, we consider $l_y^+ = \{70, 90\}$ and perform adjoint-variational data assimilation for six consecutive windows as in \S\ref{sec:ly50}.
The effect of Reynolds number will also be examined briefly.

\begin{figure}[t]
	\centering
	\includegraphics[width=\textwidth]{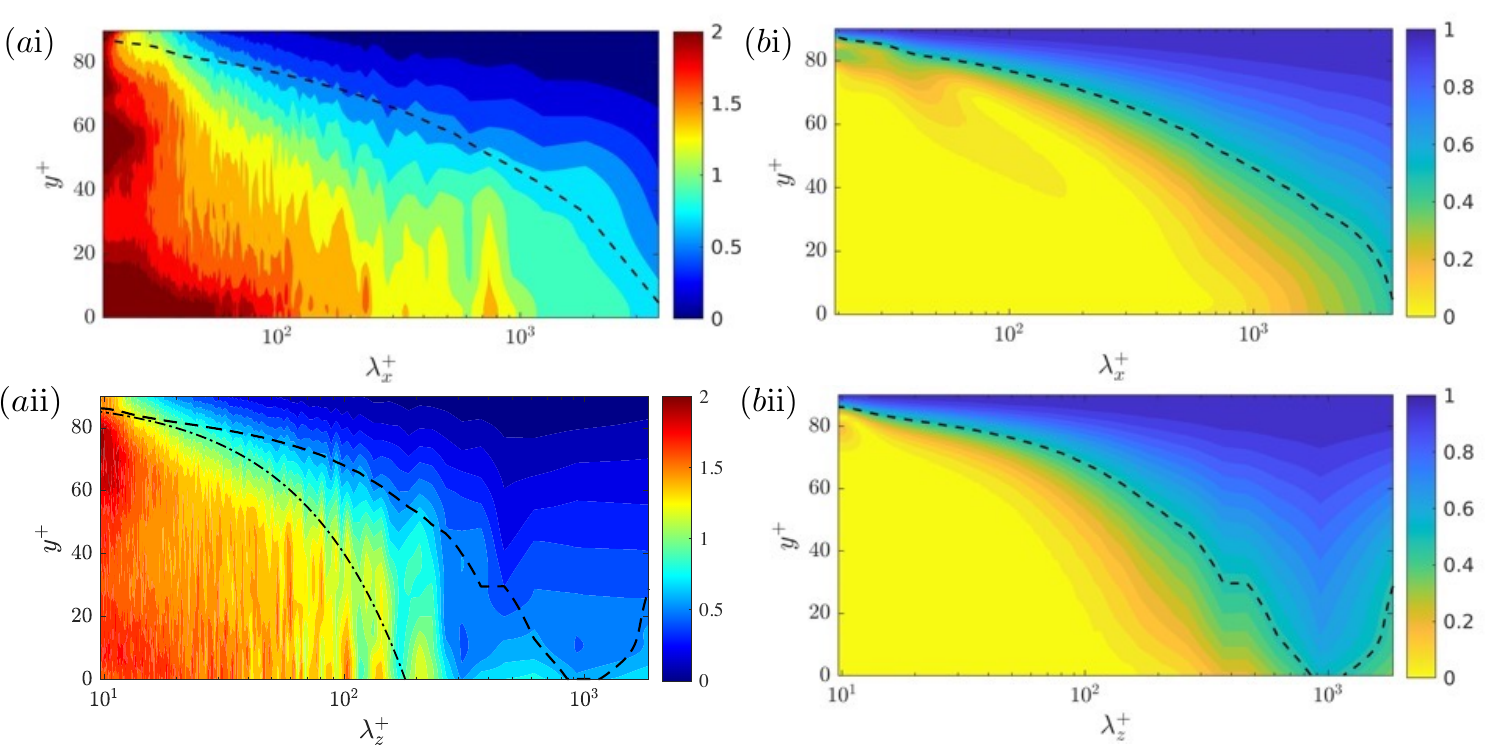}
	\caption{($a$) Wall-normal distribution of the one-dimensional Fourier spectra of estimation error of streamwise velocity, normalized by the spectra of the true state, $\mathcal{E}_{zt}(\hat{u}) / |\hat{u}_r| $. ($b$) Coherence spectra of streamwise velocity, $\mathcal{R}\left(\hat{u}(l_y),\hat{u}(y)\right)$. (i,ii) Streamwise and spanwise spectrum.
    Black dashed lines: $\mathcal{R} = 0.5$; Black dashed-dot lines in ($a$): $\lambda_z^+ = 2(l_y^+ - y^+)$. The thickness of the unknown layer is $l_y^+ = 90$.
	}
	\label{fig:ly90_spectrum_coherence_u}
\end{figure}

We first analyze the adjoint-variational reconstruction with $l_y^+ = 90$ at $Re_{\tau} = 590$.
The evolution of the cost function, estimation error, and flow statistics across the six assimilation windows are similar to figure \ref{fig:ly50_errv_cost} and thus are not repeated here.
Within the last window $t \in [5T, 6T]$, the spectra of estimation error are evaluated and visualized in figure \ref{fig:ly90_spectrum_coherence_u}$a$ for the streamwise velocity.
Within $y^+ \in [40, 90]$, the dependence of estimation error on the wavelength is similar to the $l_y^+ = 50$ case in figure \ref{fig:ly50_spectrum_uwp}$a$: longer wavelengths are reconstructed more accurately, and the associated errors are less affected by the distance from the first observation plane.
A particularly interesting region is $\{y^+ < 40, \lambda_x^+ \gtrsim 2000, \lambda_z^+ \gtrsim 200\}$, where the error of near-wall large scales becomes almost independent of the wall-normal coordinate.
However, the error of smaller scales within $y^+ < 40$ remains increasing with the distance from the observations until it saturates.
These results are reminiscent of the bimodal behaviour of the energy spectra of wall turbulence and the associated two categories of coherent structures: outer large-scale motions and inner-layer streaks \citep{hutchins2007LSM}.
The latter are predominately located at $y^+ \approx 15$ with characteristic wavelengths $\lambda_x^+ \approx 1000, \lambda_z^+ \approx 100$, which correspond to the ``inner peak'' of  the energy spectra of streamwise velocity fluctuations.
The instability of these streaks leads to the formation of quasi-streamwise vorticies and the regeneration cycle of near-wall turbulence.
Based on the numerical experiments by \cite{Jimenez1999}, the near-wall cycle is self-sustained and almost independent from turbulent motions in the outer layer.
Therefore, the inner-layer streaks are difficult to reconstruct using outer observations in $y^+ > 90$.
The other category of coherent structures, namely the outer large-scale motions, contribute to the ``outer peak'' of the of the energy spectra.
The sizes of these structures scale with outer units, $\lambda_x, \lambda_z \sim O(h)$.
Despite the residence of these structures in the outer layer, they have a significant influence on the inner-layer turbulence through the amplitude modulation mechanism, which facilitates an accurate estimation of near-wall large scales.

Although the estimation accuracy in figure \ref{fig:ly90_spectrum_coherence_u} is consistent with the physical property of coherent structures in wall turbulence, the present results should not be simply interpreted as a confirmation of previous findings.
The coherent structures are generally extracted using isosurfaces of two-point correlations that only quantify the linear dependence across different locations, while the adjoint-variational approach utilizes information from the full Navier Stokes equations.
In fact, it is illustrative to compare the spectra of estimation error with the coherence spectrum,
\begin{equation}
    \label{eq:coherence}
    \mathcal{R}\left(\hat{q}_1(\boldsymbol{k},l_y), \hat{q}_2(\boldsymbol{k},y) \right) = \frac{|\langle\hat{q}_1(\boldsymbol{k},l_y) \hat{q}_2^*(\boldsymbol{k},y)\rangle| }{ \langle |\hat{q}_1(\boldsymbol{k},l_y)|^2\rangle^{1/2} \langle |\hat{q}_2(\boldsymbol{k},y)|^2\rangle^{1/2} },
\end{equation}
which quantifies the correlation between the observation $\hat{q}_1(\boldsymbol{k},l_y)$ and the flow variable within the unknown layer $\hat{q}_2(\boldsymbol{k},y)$ in Fourier space.
To ensure convergence of the coherence spectrum, the average in (\ref{eq:coherence}) is performed over 400 advective time units.
The coherence spectra of the streamwise velocity fluctuation are plotted in figure \ref{fig:ly90_spectrum_coherence_u}$b$, and the contour lines of $\mathcal{R} = 0.5$ are reproduced in figure \ref{fig:ly90_spectrum_coherence_u}$a$ (black dashed lines) for a more direct comparison.
A key observation is that the coherence tends to underestimate the observability of near-wall turbulence from outer measurements, especially for the near-wall large scales.
For example, at $\lambda_z^+ = 300$ and $y^+ \leq 20$, the coherence is below $0.2$, suggesting a weak correlation between these structures and the outer observations, while the adjoint-variational estimation error is only $50\%$ of the true spectrum.
At an arbitrary location $y$ within the unknown layer, the accurately reconstructed spanwise waves can be better demarcated by $\lambda_z \geq 2(l_y - y)$ (dashed-dot line in panel $a$ii).
This relation is derived by assuming the structures with wavelength $\lambda_z$ centered at $y = l_y$ have a wall-normal depth proportional to $0.5 \lambda_z$.

\begin{figure}[t]
	\centering
	\includegraphics[width=\textwidth]{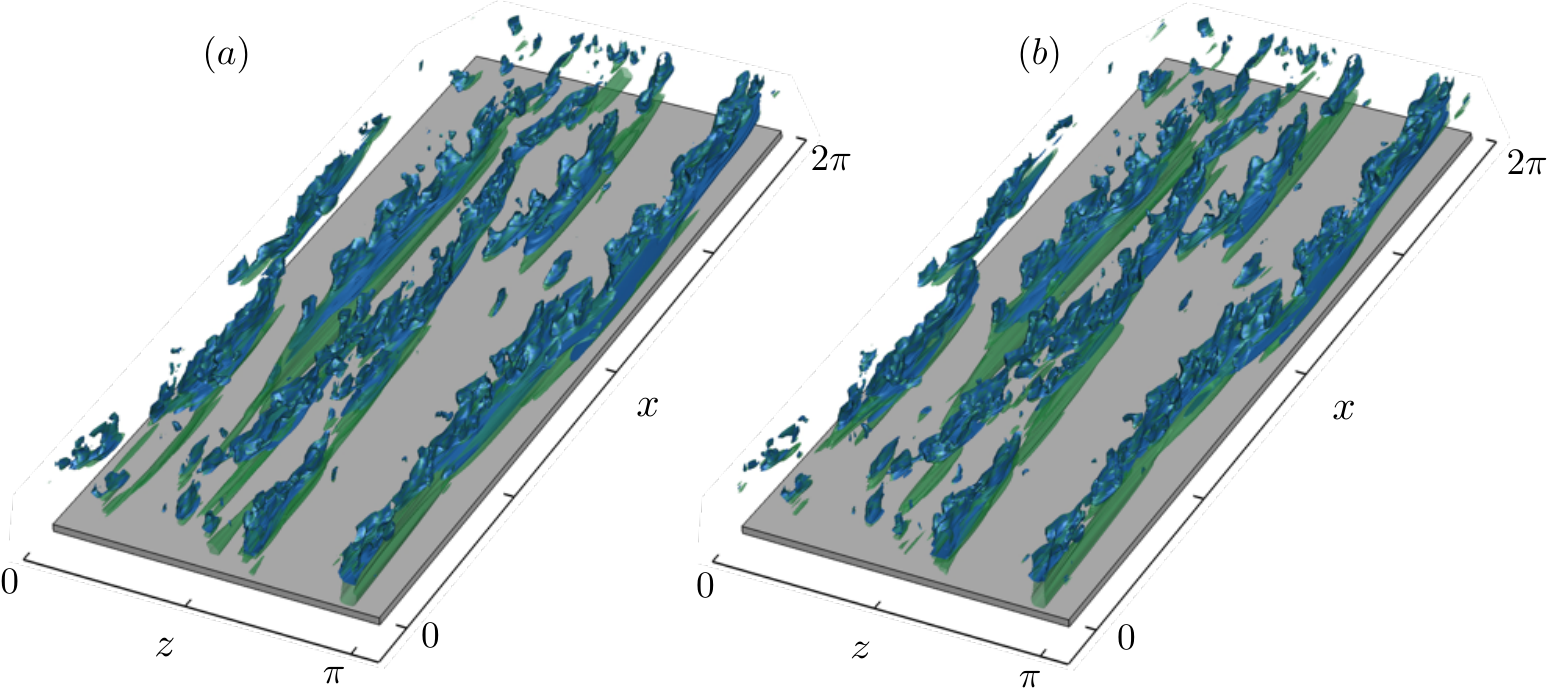}
	\caption{Filtered streamwise velocity fluctuation of ($a$) adjoint-variational estimation and ($b$) the true state, visualized by isosurfaces of $u^{\prime} = -0.12$ within $y^+ \in [0,90]$.
    Blue isosurfaces: the filtered wavelengths satisfy $\mathcal{R}\left(\hat{u}(\lambda_x,l_y),\hat{u}(\lambda_x,y)\right) \geq 0.5$ and $\mathcal{R}\left(\hat{u}(\lambda_z,l_y),\hat{u}(\lambda_z,y)\right) \geq 0.5$.
    Green isosurfaces: same filter for $\lambda_x$ as blue, but $\lambda_z^+ \geq 2(l_y^+ - y^+) $.
	}
	\label{fig:ly90_3d_u}
\end{figure}

The predictions of near-wall structures from outer measurements using the adjoint approach and based on properties of the coherence spectra are further differentiated in figure \ref{fig:ly90_3d_u}.
The blue isosurfaces represent the streamwise fluctuations extracted using a coherence-based filter, where the wavelengths $(\lambda_x,\lambda_z)$ of the filtered field satisfy $\mathcal{R}\left(\hat{u}(\lambda_x,l_y),\hat{u}(\lambda_x,y)\right) \geq 0.5$ and $\mathcal{R}\left(\hat{u}(\lambda_z,l_y),\hat{u}(\lambda_z,y)\right) \geq 0.5$.
The same streamwise filter is adopted for the green isosurfaces, but the spanwise filter is changed to $\lambda_z \geq 2(l_y - y) + 2\Delta z$.
In the vicinity of observed outer layer, both filters capture a wide range of scales in the adjoint-variational estimation (panel $a$) that match the true state (panel $b$).
Deeper towards the wall, the coherence-based filter fails to identify the majority of the large-scale structures that are accurately reconstructed by the adjoint approach.
Another advantage of adjoint-variational data assimilation is the quantification of causality among different flow structures, or events, in turbulence.
Only the structures that influence the observed signals can be accurately reconstructed.
In other words, the flow structures in the unknown layer can be classified as ``observation-attached'' and ``observation-detached'' eddies, although the term ``attached" takes on a more general meaning since it entails an influence on the observations within the observation horizon, and through the dynamics of the Navier-Stokes equations.
The ``observation-attached" eddies in the present case consist of the small-scale motions near the observed region and the large scales that penetrate across the wall layer.
These structures align with the domain of dependence of observations, which is determined by marching the adjoint equations backward in time \citep{Wang2022Hessian}.
The ``observation-detached" eddies are primarily near-wall structures, ranging from self-sustained streaks to Kolmogorov eddies that are not encoded in the outer observations.

\begin{figure}[t]
	\centering
	\includegraphics[width=\textwidth]{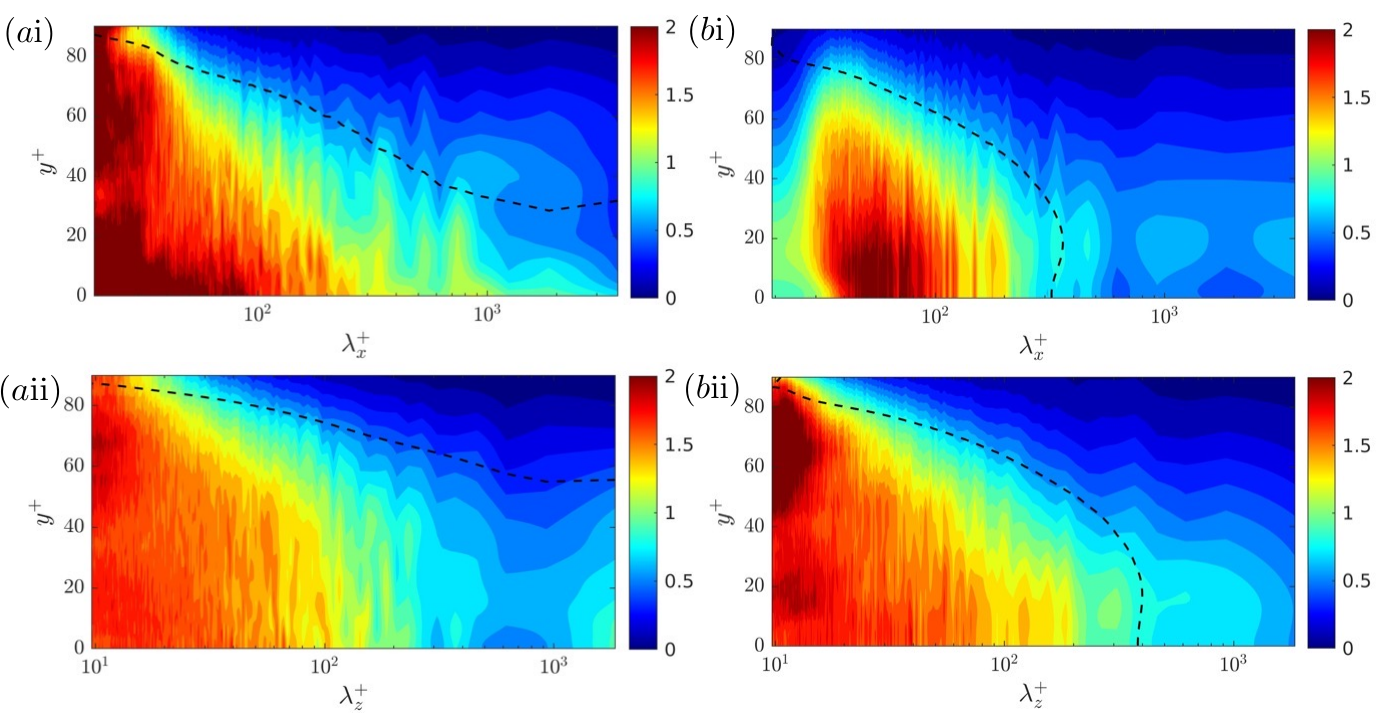}
	\caption{One-dimensional Fourier spectra of the estimation error of ($a$) spanwise velocity and ($b$) pressure, evaluated at different wall-normal locations and normalized by the spectrum of the respective true field.
    (i,ii) Streamwise and spanwise spectra.
	}
	\label{fig:ly90_spectrum_wp}
\end{figure}

The spectra of estimation error of the spanwise velocity, shown in figure \ref{fig:ly90_spectrum_wp}$a$, are analogous to the streamwise velocity in figure \ref{fig:ly90_spectrum_coherence_u}$a$.
The quasi-streamwise rolls are reconstructed with the highest accuracy.
The discrepancy between the adjoint estimation error and the coherence spectra ($\mathcal{R}(\hat{w},\hat{w})=0.5$, black dashed lines) is more evident here than in figure \ref{fig:ly90_spectrum_coherence_u}$a$.
The coherent structures identified using $\mathcal{R}(\hat{w},\hat{w})=0.5$ never touch the wall, while the adjoint approach successfully reconstructs the large-scale $w^{\prime}$ structures that are attached to the wall.
The comparison between error spectra of pressure and the coherence spectra $\mathcal{R}(\hat{p},\hat{p})=0.5$ is provided in figure \ref{fig:ly90_spectrum_wp}$b$.
As reported in previous studies \citep{sillero2014twopoint}, the coherence-based pressure structures are moderately elongated along the spanwise direction.
An accurate reconstruction of these structures can be difficult since the outer pressure data are not available for the adjoint approach.
Due to the non-local property of pressure, the outer pressure field, especially near the edge of observed region $y = l_y$, cannot be fully determined from the velocity measurements.
Nevertheless, the adjoint approach effectively captures the dominant coherent structures of pressure in the near-wall layer.

\begin{figure}[t]
	\centering
	\includegraphics[width=\textwidth]{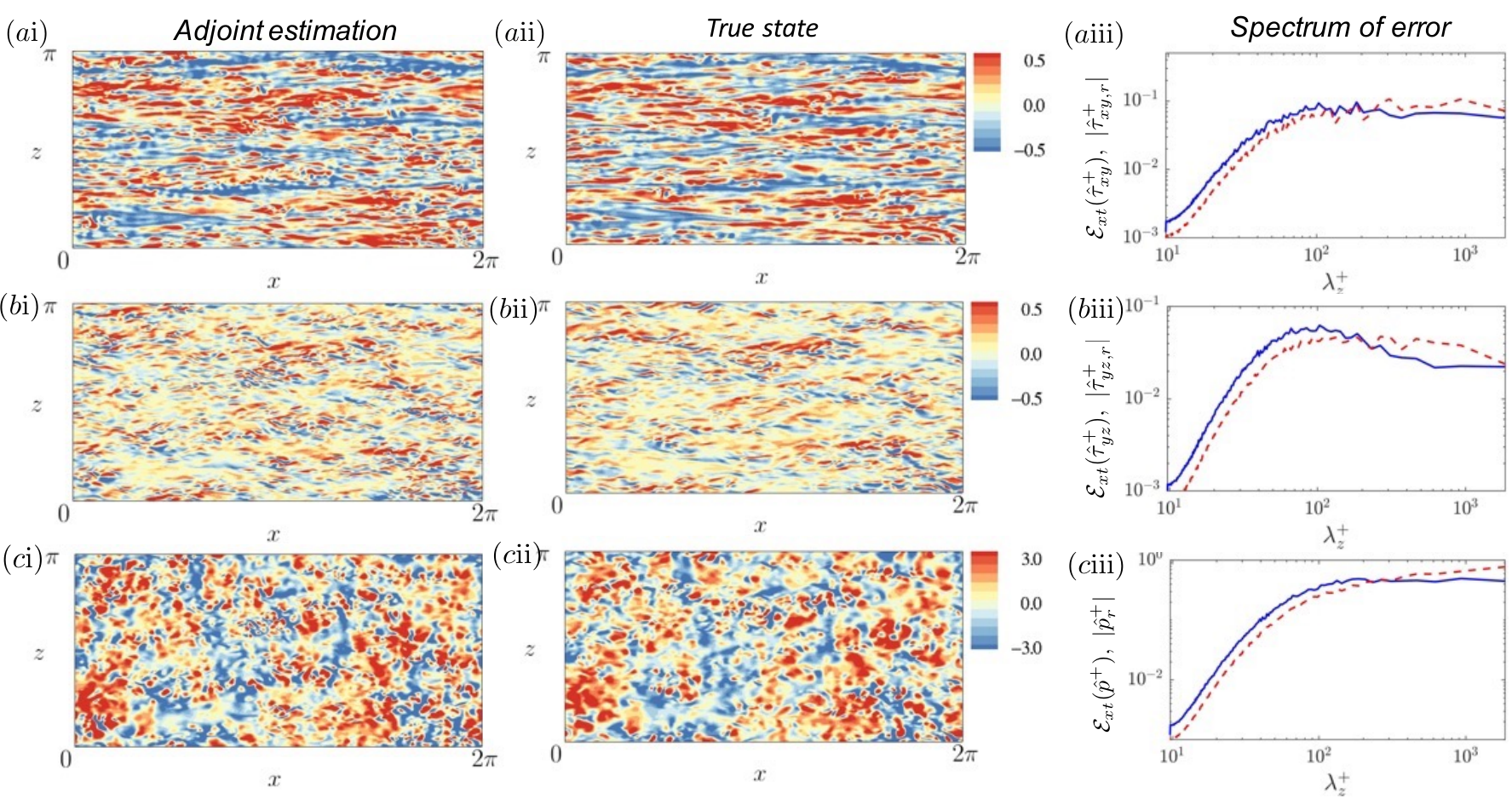}
	\caption{(i) Adjoint-variational estimation of fluctuations of wall shear stress and pressure at $t = 6T$ when $l_y^+ = 90$. (ii) The true fields. (iii) (Blue) spanwise spectra of the estimation error of wall quantities averaged in $t \in [5T,6T]$, compared with (red dashed) the spectra of the true field. ($a$,$b$,$c$) two components of the wall shear stress and pressure, $\{\tau_{xy}^+, \tau_{yz}^+,p^+\}$.
	}
	\label{fig:ly90_vis_wall_spectrum}
\end{figure}

\begin{figure}[t]
	\centering
	\includegraphics[width=\textwidth]{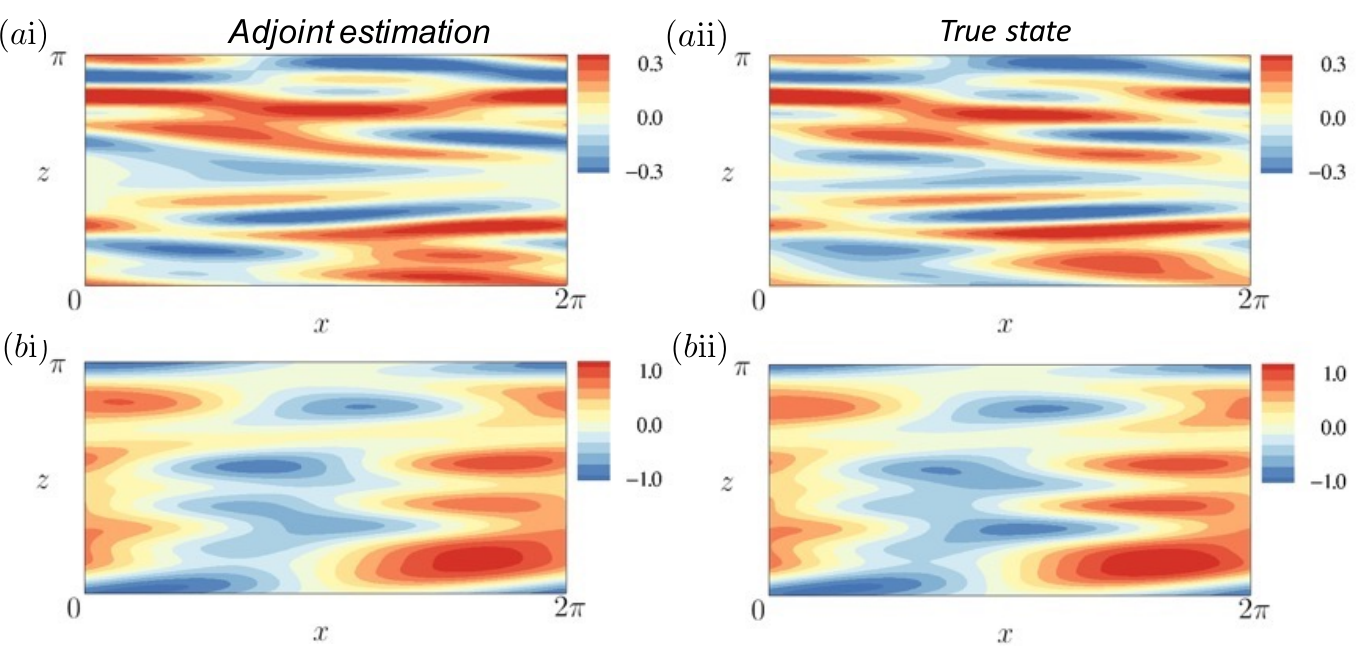}
	\caption{Large-scale structures of the ($a$) streamwise wall shear stress ($b$) wall pressure, visualized using (i) adjoint-variational estimation and (ii) the true state. Large scales are extracted by applying a low-pass filter ($\lambda_x^+ \geq 3000, \lambda_z^+ \geq 180$) to the full field.
	}
	\label{fig:ly90_vis_filtered}
\end{figure}

The adjoint reconstructions of the wall shear stresses and pressure are visualized in figures \ref{fig:ly90_vis_wall_spectrum}(i), and compared with the true fields in panels (ii).
Overall the estimation is less accurate than in the $l_y^+ = 50$ case reported in figure \ref{fig:ly50_vis_wall_spectrum}.
Here, most of the streaks and local extrema of stresses are captured, but their locations and magnitudes are not precisely reproduced.
The estimation errors in Fourier space are quantified in figure \ref{fig:ly90_vis_wall_spectrum}(iii).
The errors in small scales are approximately $\sqrt{2}$ times the true spectra of the corresponding wall signal, which indicates that the reconstructed dissipative eddies are nearly independent from the true state.
For longer waves, the errors are lower than the true spectra, which is consistent with the qualitative description in panels (i) and (ii).
These structures can also be extracted using the same low-pass filter as in figure \ref{fig:ly90_3d_u} for the green isosurfaces.
At the wall, the filtered wavelengths, $\lambda_x^+ \geq 3000$ and $\lambda_z^+ \geq 180$, approximately coincide with the intersections between the spectra of error and the true state in figure \ref{fig:ly90_vis_wall_spectrum}(iii).
The filtered structures of the streamwise wall shear stress and pressure are visualized in figure \ref{fig:ly90_vis_filtered}.
The locations and shapes of the estimated large scales are similar to the true filtered fields, which demonstrates that accurate prediction of the near-wall large scales is essential to reproduce the outer large-scale observations.

\begin{figure}[t]
	\centering
	\includegraphics[width=\textwidth]{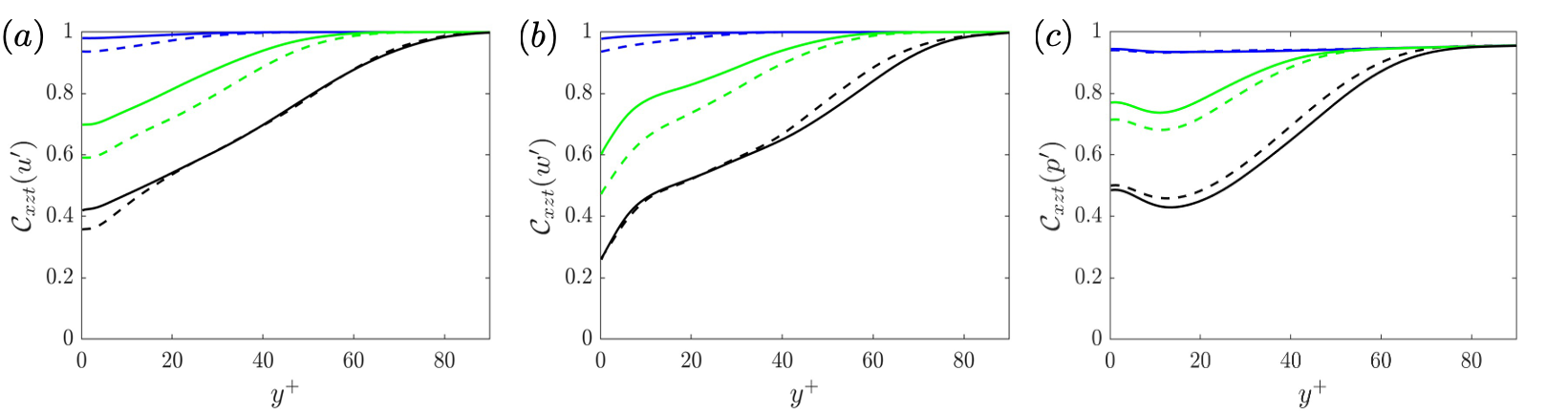}
	\caption{Influence of thickness of the unknown layer and Reynolds number on the estimation accuracy, quantified using the correlation coefficient between the estimated and true state. The correlations are averaged along the horizontal directions and in time, $t \in [5T,6T]$. ($a$,$b$,$c$) Fluctuations of streamwise velocity, spanwise velocity, and pressure. (Blue, green, black) $l_y^+ = \{50,70,90\}$. (Dashed, solid) $Re_{\tau}=\{392,590\}$.
	}
	\label{fig:summary_ly_Retau}
\end{figure}

Figure \ref{fig:summary_ly_Retau} illustrates the impact of the thickness of the unknown layer $l_y$ and of the Reynolds number on the estimation accuracy.  The figure reports the correlation coefficient between true and estimated states,
\begin{equation}
    \label{eq:cc}
    \mathcal{C}_{xzt}(q) = \frac{\langle q q_r\rangle_{xzt}}{\langle q^2\rangle_{xzt}^{1/2} \langle q_r^2\rangle_{xzt}^{1/2}}.
\end{equation}
The average in equation (\ref{eq:cc}) is performed along the horizontal directions and over the last assimilation window $t \in [5T,6T]$.
Within the observed region ($y > l_y$), the correlations of the streamwise and spanwise fluctuations are equal to unity since the adjoint approach accurately reproduces the velocity observations, and the correlation of pressure remains higher than $0.95$ despite the lack of pressure data.
Outside the observed region, the correlation remains close to unity when $l_y^+ = 50$.
As the unknown layer is expanded ($l_y^+ = 70,90$), the correlation decays with distance from the first observation plane towards the wall.
Note that, when the profiles of these correlations are plotted versus $y^+$, they appear approximately independent of $Re_{\tau}$.  The trend indicates that the decay in the correlations is primarily due to the inaccurate estimation of the inner-layer streaks and smaller scales.
These structures dominate the kinetic energy of the near-wall turbulence, and their sizes scale with viscous units.

\begin{figure}[t]
	\centering
	\includegraphics[width=0.5\textwidth]{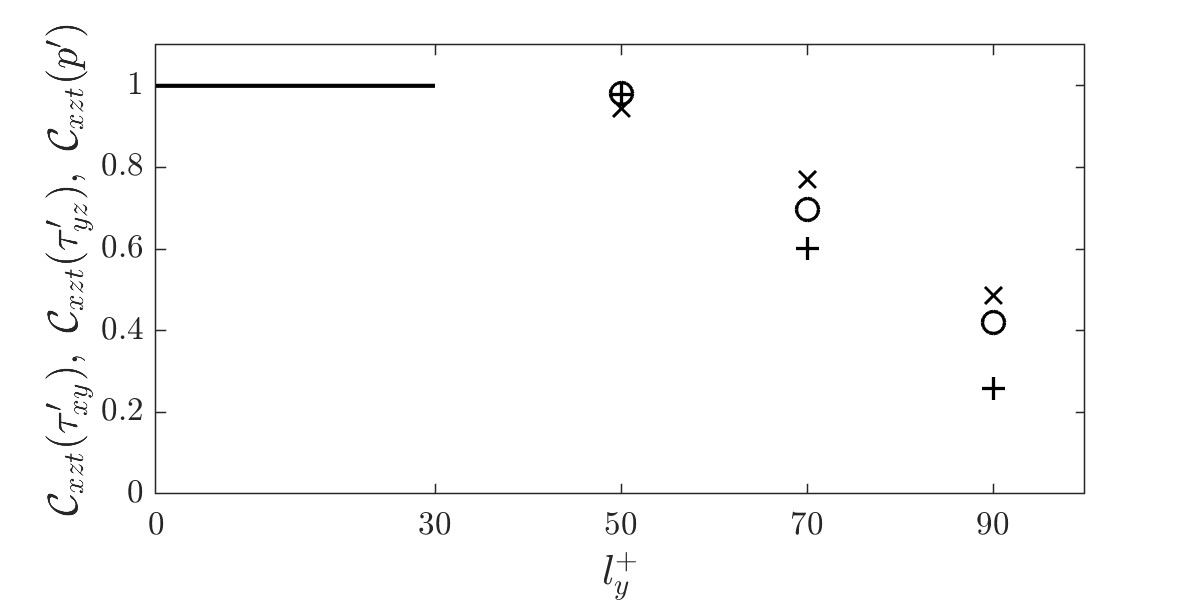}
	\caption{Estimation accuracy of wall signals versus thickness of the unknown layer, when fully resolved outer observations are adopted.
    The accuracy is quantified using the correlation coefficient between the true wall signals and their estimations. Symbols ($\circ, +, \times$) correspond to the two components of wall shear stress and pressure ($\tau_{xy},\tau_{yz},p$).
    A version of this figure featured in the recent review by \citet[][CC BY 4.0]{zaki2025}.
	}
	\label{fig:cc_wall_ly}
\end{figure}

The highest attainable prediction accuracy of the wall stresses and pressure, given fully resolved outer observations, is summarized in figure \ref{fig:cc_wall_ly}.
When $l_y^+ \lesssim 30$, we can achieve a perfect synchronization of near-wall turbulence by continuously injecting observations into the Navier-Stokes solution \citep{Wang2022synch}.
Therefore, the correlation coefficient between the true and estimated quantities is precisely equal to unity.
Beyond thirty viscous units, synchronization is no longer feasible.
However, the wall stresses and pressure can still be accurately reconstructed for $l_y^+ \leq 50$ using the adjoint-variational approach. 
As the observations are farther separated from the wall, the correlation coefficient monotonically decreases, reaching $\left(\mathcal{C}(\tau_{xy}^{\prime}), \mathcal{C}(\tau_{yz}^{\prime}), \mathcal{C}(p^{\prime}) \right) = (0.42, 0.26, 0.49)$ at $l_y^+ = 90$.
Unlike the linear state estimation methods, the adjoint reconstruction of wall stresses and pressure are not only correlated with the outer flow, but also dynamically and causally consistent with the Navier-Stokes evolution reproducing the outer observations.

\subsection{The effect of measurement resolution and filtering}
\label{sec:filtered_data}

The discussion thus far has assumed access to fully resolved velocity data in the outer layer, which is difficult to acquire in practice.
It is also desirable to investigate whether the estimation accuracy achieved in the previous sections is robust against limited measurement resolution in the outer layer.
Intuitively, if the data resolution is in the limit of excessively coarse data resolution, the estimation quality of the wall layer will be significantly compromised.
In this section, we consider what can be regarded as moderately coarse velocity data in the outer layer, and focus on the estimation accuracy of wall stresses and pressure at $Re_{\tau} = 590$.

The outer velocity fields that furnish the observations is first filtered and then sub-sampled.
Specifically, the fully resolved fields obtained from DNS are box-filtered along the horizontal directions,
\begin{equation}
    \label{eq:filter}
    \tilde {\boldsymbol u}(\boldsymbol x,t) = \int G(\boldsymbol r) \boldsymbol u(\boldsymbol x - \boldsymbol r,t) d \boldsymbol r.
\end{equation}
The filter function $G(\boldsymbol{r})$ is defined as
\begin{equation}
    \label{eq:box_filter}
    G(\boldsymbol r) = G(r_x,r_z) = \frac{1}{\Delta_{fx}\Delta_{fz}} H\left(\frac 12 \Delta_{fx} - |r_x| \right) H\left(\frac 12 \Delta_{fz} - |r_z|\right),
\end{equation}
where $H(\cdot)$ is the Heaviside step function and $(\Delta_{fx},\Delta_{fz})$ are filter widths in the streamwise and spanwise directions.
The filtered velocity is then sub-sampled on every eighth grid point in each dimension, including space and time.
As a result, the sparse observations are separated by $\Delta x_m^+ = 77$, $\Delta y_m^+ \in [12,52]$, and $\Delta z_m^+ = 39$, and their temporal resolution is $\delta t_m^+ = 0.51$ , which are comparable to typical resolutions of experimental measurements of wall turbulence \citep{lu2024scaling}.
Although the Kolmogorov eddies are not resolved by the observations, such a setup has been demonstrated sufficient to accurately reconstruct the turbulence in between the observation sites, when observations are available throughout the channel volume \citep{Wang2021}, while here in contrast we only consider outer observations.  
The filter width is selected slightly smaller than the observation spacing, $\Delta_{fx}^+ = 58$ and $\Delta_{fz}^+ = 29$, which correspond to the size of six DNS grid cells in $x$ and $z$ respectively, $\Delta_f / \Delta_{DNS} = 6$.
The observation operator $\mathcal{M}$ can therefore  be expressed as $\mathcal{M}(\boldsymbol{u}(\boldsymbol{x})) = \tilde{\boldsymbol{u}}(\boldsymbol{x})|_{\boldsymbol{x} = \boldsymbol{x}_m}$.
To separate the influence of filtering and sub-sampling, we also consider a case where the observations are not filtered before sub-sampling, denoted as $\Delta_f / \Delta_{DNS} = 0$.

\begin{figure}[t]
	\centering
	\includegraphics[width=\textwidth]{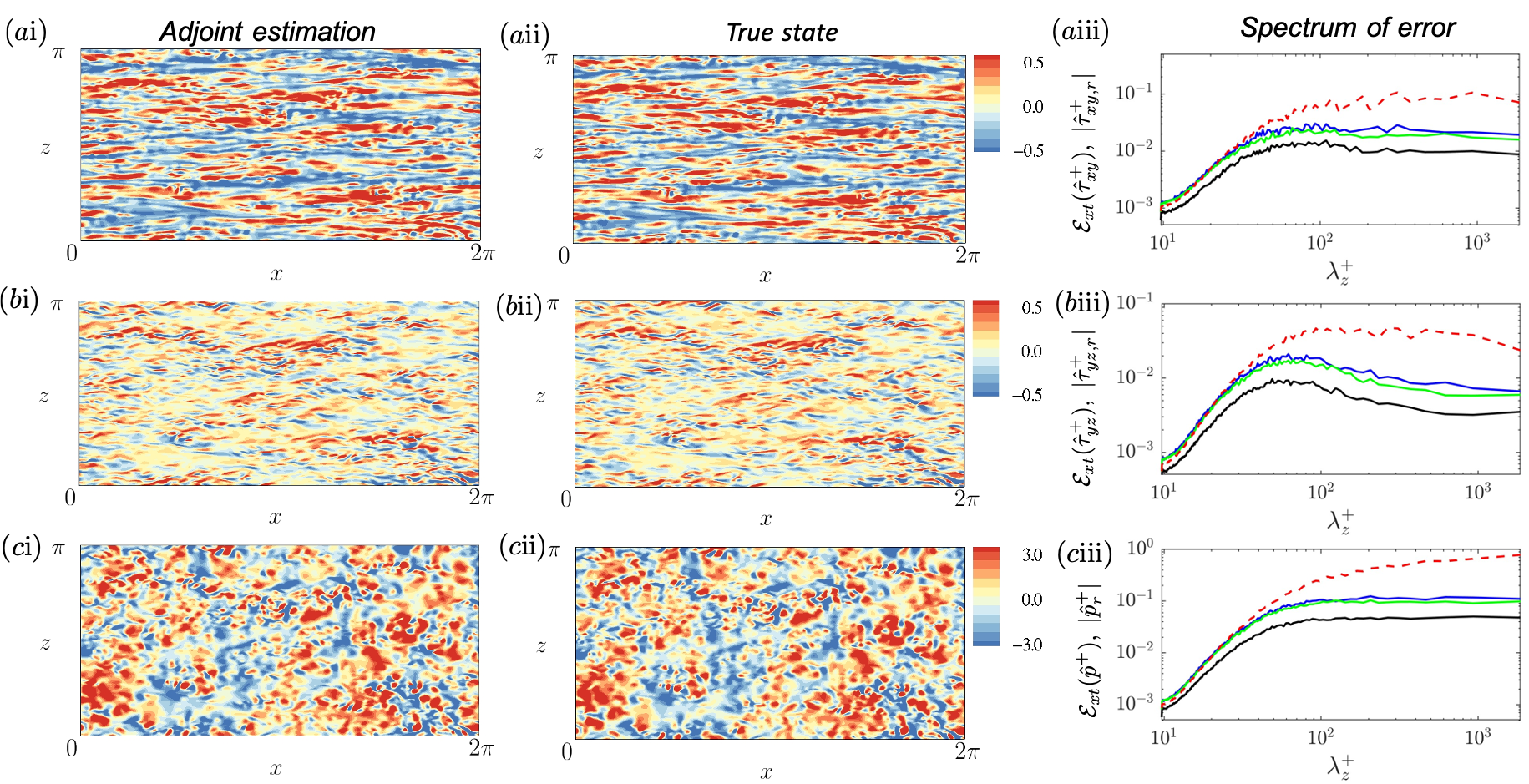}
	\caption{(i) Adjoint estimation of wall signals at $t = 6T$ using filtered and sub-sampled outer observations beyond $l_y^+ = 50$. (ii) The true fields.(iii) Spectra of the estimation error when observations are (black) fully resolved outer data, (blue) sub-sampled data, $\Delta_f / \Delta_{DNS} = 0$, or (green) filtered and sub-sampled data $\Delta_f / \Delta_{DNS} = 6$.
    (Red) spectra of the true fields.
    ($a$,$b$,$c$) two components of the wall shear stress and pressure, $\{\tau_{xy}^+, \tau_{yz}^+,p^+\}$.
	}
	\label{fig:ly50filter_vis_wall_spectrum}
\end{figure}

The data assimilation procedures and the sliding-window strategy are almost the same as in \S\ref{sec:ly50} and \S\ref{sec:ly90}, with only two differences.
Firstly, due to the limited data resolution in the outer layer, the adjoint-variational algorithm is now applied to reconstruct the full initial state, not just the wall layer.
The other difference is the first guess of the initial state.
Linear stochastic estimation is still adopted to approximate the initial velocity field in the wall layer, but only the coarse observations are fed into LSE, not the fully resolved data.
Within the observed outer layer, the initial velocity is estimated between observation locations using spline interpolation of the sparse velocity data.

Given filtered and sub-sampled data with $l_y^+ = 50$, the adjoint estimation of the wall stresses and pressure at $t = 6T$ are visualized in figure \ref{fig:ly50filter_vis_wall_spectrum}(i).
Compared with the true fields in panel (ii), only a few differences in the small scales can be identified, and the estimation quality is comparable to the results using fully resolved observations in figure \ref{fig:ly50_vis_wall_spectrum}.
Due to the small differences in these qualitative comparisons, it is necessary to perform a more detailed analysis of the influence of data resolution by examining the errors in spectral space, as shown in panels (iii).
The reconstruction is slightly more accurate when the outer observations are filtered before the sub-sampling (from blue to green lines), because the filtering operation incorporates more information about the flow field.
Regardless of the filter width, the error of adjoint estimation using under-resolved outer observations (blue, green lines) are higher than the results using fully resolved data (black lines) across all the scales.
This trend can be explained from the assimilation perspective:
as the outer observations become coarser, they can be traced back to a larger set of turbulent states, therefore increasing the fundamental difficulty of searching for the true turbulent state.
Nevertheless, the estimation error using limited outer observations remains an order of magnitude smaller than the true spectra at large scales, which aligns with the high estimation quality in panels (i).
The reconstructed smaller scales are still correlated with the true state, as evidenced by the estimation error being lower than than $\sqrt{2}$ times the true spectra.
In summary, the estimation accuracy of wall signals is compromised but not significantly affected by the decreased resolution of the data.

\begin{figure}[t]
	\centering
	\includegraphics[width=\textwidth]{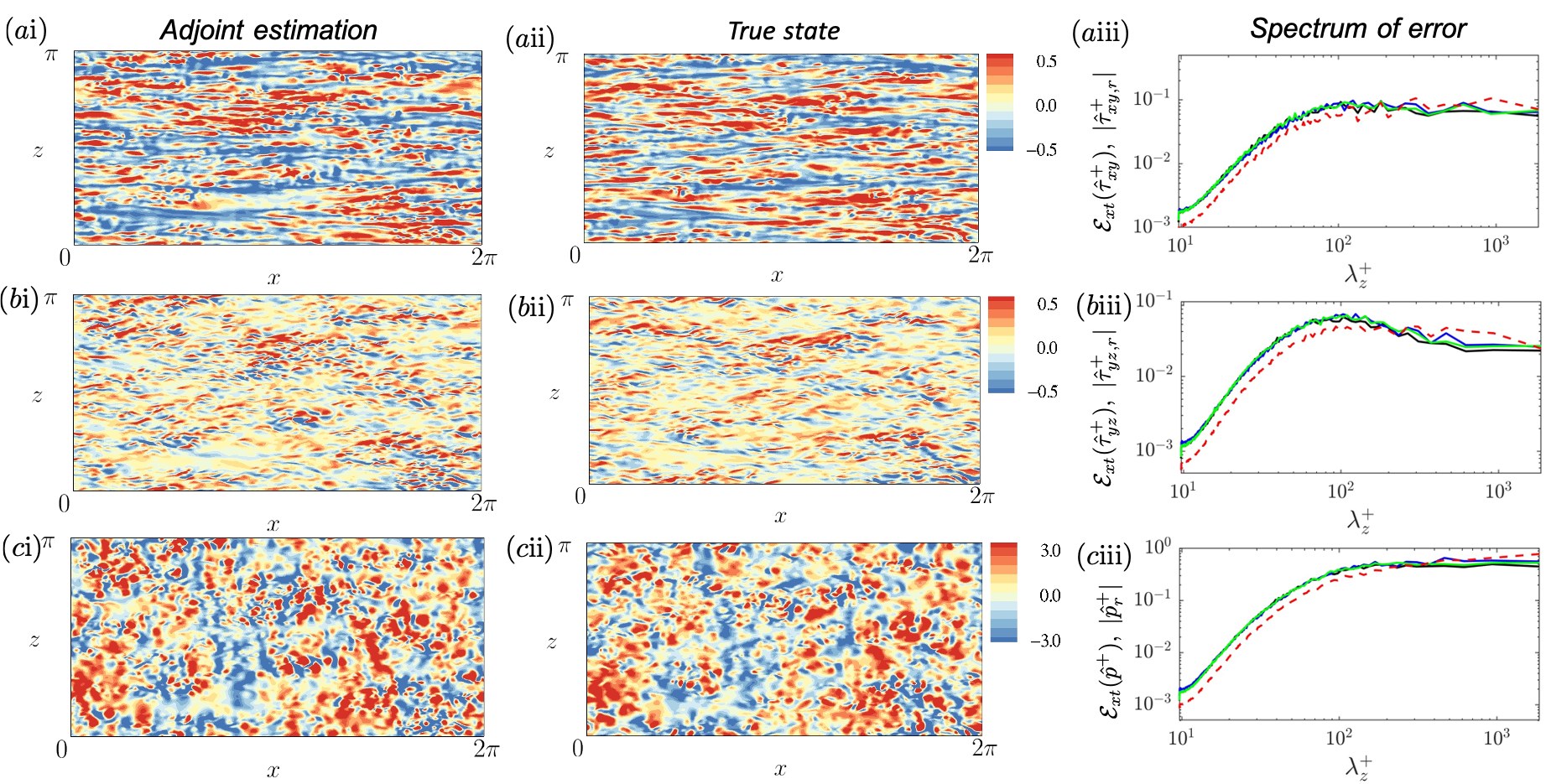}
	\caption{Same as figure \ref{fig:ly50filter_vis_wall_spectrum} but the thickness of the unknown layer is $l_y^+ = 90$. 
	}
	\label{fig:ly90filter_vis_wall_spectrum}
\end{figure}

The robustness of the estimation accuracy against measurement resolution is further verified at $l_y^+ = 90$ in figure \ref{fig:ly90filter_vis_wall_spectrum}.
In panels (i), the energy-containing large-scale structures reconstructed using under-resolved data closely resemble the true state in panels (ii), although discrepancies are evident at other scales.
The estimation error in Fourier space is largely unaffected by the measurement resolution or filtering, as illustrated in panels (iii), with only a slight increase in error at large scales when using coarser observations.
These results indicate that as observations move further from the wall, the decoded wall signals remain predominantly sensitive to the large scales in the outer layer, which are well captured by the observations despite the coarser data resolution.

\begin{figure}[t]
	\centering
	\includegraphics[width=0.5\textwidth]{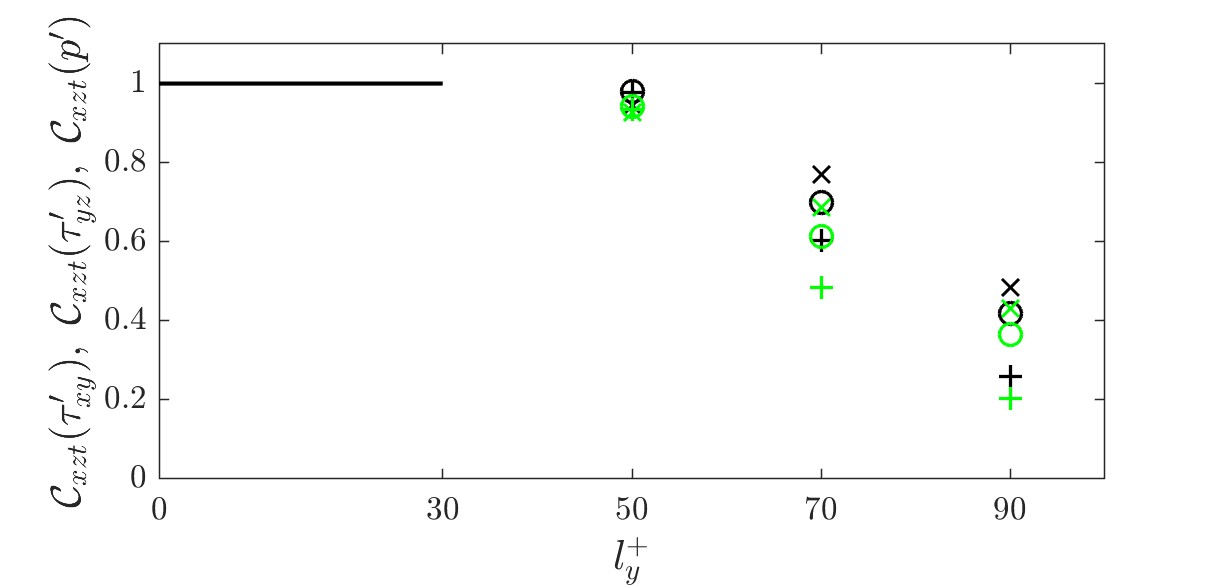}
	\caption{Estimation accuracy of wall signals versus thickness of the unknown layer. (Black) estimation using fully resolved outer data; (green) estimation using filtered and sub-sampled data.
    The accuracy is quantified by the correlation coefficient between the true wall signals and their estimations.
    Symbols ($\circ, +, \times$) correspond to the two components of wall shear stress and pressure ($\tau_{xy},\tau_{yz},p$).
	}
	\label{fig:cc_wall_ly_filter}
\end{figure}

Combining the estimation accuracy using filtered and sub-sampled data with figure \ref{fig:cc_wall_ly}, we summarize the influence of measurement resolution and amount of observations on the highest attainable accuracy of wall signals in figure \ref{fig:cc_wall_ly_filter}.
The estimated wall stresses and pressure remain accurate up to $l_y^+ = 50$, with a correlation coefficient exceeding 0.95.
This high level of accuracy is maintained for the herein considered sub-sampled / under-resolved outer observations, whose density is 1/4096 of the requirement for DNS.
Beyond $l_y^+ = 50$, the estimation accuracy diminishes with increasing thickness of the unknown layer, yet remains robust against moderate coarse-graining of the outer observations.
Taken altogether, these results underscore the potential for an accurate estimation of wall signals even with reduced measurement resolutions in the outer layer, highlighting the effectiveness of the adjoint-variational approach that strongly enforces the full Navier-Stokes equations.

\section{Conclusions}
\label{sec:conclusions}

Using adjoint-variational data assimilation, we sought an optimal reconstruction of near-wall turbulence from outer observations.
Under the constraint that the evolution of the estimated flow state exactly satisfies the Navier-Stokes equations, we searched for the initial condition that minimizes a cost function defined as the difference between a time series of observations and their estimation.
By solving the forward equations and their discrete adjoint, the gradient of the cost function was evaluated and utilized in the L-BFGS optimization algorithm to iteratively update the estimated initial state.
To mitigate the influence of the chaotic nature of turbulence, a sliding-window strategy was adopted.  Specifically, the observation time horizon was divided into consecutive assimilation windows, each shorter than the Lyapunov timescale.
Data assimilation was performed within each window, and then sequentially from one window to the next.  At the end of the total assimilation time horizon, the estimated flow state was statistically stationary while also optimally reproducing the available outer observations.

The performance of the adjoint-variational data assimilation was examined in detail for an unknown wall-attached layer with thickness $l_y^+ = 50$, at friction Reynolds number $Re_\tau=590$.  
After six assimilation windows ($6T^+ = 183$), the estimated state was statistically stationary, and accurately shadowed the true state as evidenced by reproducing the observations in the outer layer.
In the unknown wall layer, the estimated flow statistics, including r.m.s.~fluctuations and probability distributions of wall stress and pressure, are identical to the true state.
The estimation error was 77\% smaller than linear stochastic estimation using contemporaneous outer observations.
Spectral analysis revealed that the estimated near-wall turbulence remained correlated with the true state across all scales, from the largest energy-containing motions to the smallest dissipative Kolmogorov eddies.
Furthermore, the error of large scales were an order of magnitude smaller than the true spectra of the flow variables throughout the entire unknown layer.

As the first observation plane is placed farther from the wall, an accurate estimation of all the scales of near-wall turbulence becomes more difficult.
Up to $l_y^+ = 90$, the reconstructed large scales remain accurate regardless of the wall-normal location, whereas the estimation quality of smaller scales significantly deteriorates with increasing distance from observations.
These trends are consistent with the coherence spectra between outer observations and inner unknown variables, which dictates the accuracy bound of linear estimation methods.
However, the coherence spectra tend to underestimate the observability of large scales close to the wall as predicted by the adjoint assimilation.
The adjoint approach enables a rigorous quantification of the observability of near-wall turbulence from outer data, through the dynamics and the causal connection enforced by the governing equations.  
Only the observation-attached eddies, defined as the structures that influence the outer flow within the assimilated horizon, can be accurately reconstructed.
The observation-detached eddies, including inner-layer streaks and Kolmogorov scales, are self-sustained and relatively independent of the outer motions.

The dependence of the optimal estimation accuracy on the thickness of the unknown layer and Reynolds number was systematically analyzed.
When $l_y^+ \leq 30$, the upper bound of the correlation coefficient between the estimated and true states is precisely unity, due to the perfect synchronization of near-wall turbulence to the outer flow.
Beyond the threshold for synchronization, the optimal prediction accuracy of wall stresses and pressure remains close to unity when the first observation plane is up to $l_y^+ = 50$, and then gradually decays to on the order of $0.4$ at $l_y^+ = 90$.  At the two Reynolds numbers considered, $Re_\tau =\{392, 590\}$, the accuracy of the estimation was generally unchanged.

Motivated by practical interests, our final tests focused on reconstructing near-wall turbulence from under-resolved observations in the outer layer.
The observations were filtered and sub-sampled at 1/4096 of the spatio-temporal resolution required for DNS.
Compared with the optimal prediction using fully resolved observations, the estimation accuracy is mildly compromised by the coarser resolutions of the data.
Nevertheless, the estimated near-wall large-scale motions closely resemble the true state, even for $l_y^+ = 90$.
The accuracy of the wall shear stresses and pressure were verified to be robust against the reduction in the resolution of the outer observations.
The present results demonstrate the potential of augmenting practical outer measurements to predict the instantaneous wall signals with satisfactory accuracy.
If the experimental measurements are not distributed throughout the entire outer layer but confined in a small volume, such as tomographic PIV, the forward and adjoint Navier-Stokes equations can be solved exclusively within the measurement volume to reduce the computational cost. Both the initial condition and time-dependent boundary conditions of the volume will be reconstructed using the adjoint-variational algorithm. This approach is being actively explored with real experimental data \citep{lu2024scaling}.

A worthy extension of this work would be to explore the behaviour at high Reynolds numbers, where the outer large-scale motions become more energetic and have an appreciable impact on the near-wall region.  It is important to underscore, however, that it is the dual to this influence that matters for the purpose of the assimilation, specifically which near-wall scales of the turbulence that must be correctly predicted to preserve an accurate evolution of the observed outer structures.

\par\vspace*{4pt}   
\noindent
\textbf{Funding.} This work was supported by the Office of Naval Research (N00014-20-1-2715). 

\par\vspace*{4pt}   
\noindent
\textbf{Declaration of interests.} 
The authors report no conflict of interest.

\par\vspace*{4pt}   
\noindent
\textbf{Author ORCIDs.} \\
Mengze Wang, \url{https://orcid.org/0000-0002-5417-7958};\\
Tamer A. Zaki, \url{https://orcid.org/0000-0002-1979-7748}.

\appendix

\section{Influence of sliding-window size on estimation accuracy}
\label{sec:window}

\begin{figure}[t]
    \centering
    \includegraphics[width=0.9\textwidth]{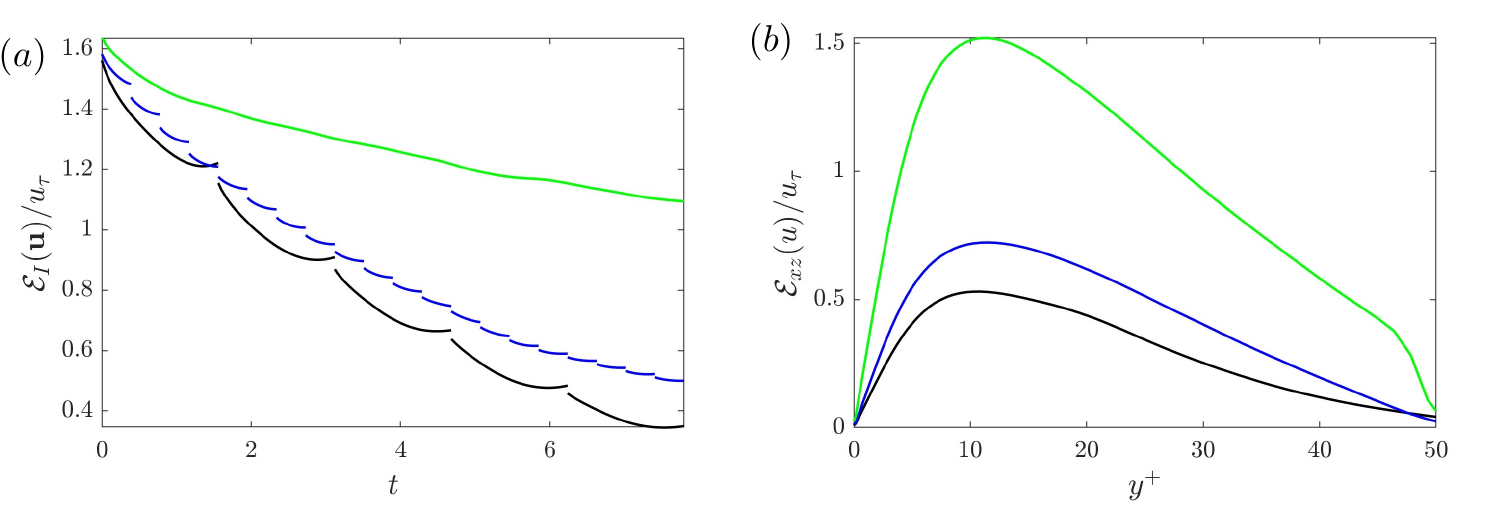}
    \caption{The effect of the duration of each assimilation window $T$ on the estimation accuracy. ($a$) The volume-averaged error in the predicted velocity versus time. ($b$) Horizontally-averaged error of streamwise velocity at time $t = 7.8$ ($t^+ = 174$), for (green, blue, black) $T^+ = \{0.87,8.7,35\}$.
    }
    \label{fig:ly50_effect_T}
\end{figure}

In \S\ref{sec:adjoint}, the sliding window size $T$ was designed to be on the order of the Lyapunov timescale, specifically we set $T = 0.8\tau_{\sigma}$ for all data-assimilation cases presented in the main text. 
In this appendix, we investigate the impact of the window size on the accuracy of estimating the near-wall turbulence.
Our analysis focuses on $Re_{\tau} = 392$, when the thickness of the unknown layer is $l_y^+ = 50$.
Two window sizes are considered, $T^+ = \{0.87, 8.7\}$, which are respectively $1/40$ and $1/4$ of the value $T^+ = 35$ used in \S\ref{sec:results}.
Within each window, we always perform 100 forward-adjoint loops to make a fair comparison.
The results are summarized in figure \ref{fig:ly50_effect_T}.
Starting from the same first guess at $t = 0$, a shorter window leads to slower decay of the estimation error (panel $a$).
Therefore, if the sliding-window data assimilation is stopped at the same time ($t=7.8$, or $t^+ = 174$), the lowest estimation error is achieved with the longest window size.
In addition, there is an abrupt change in the state at the boundary between windows, which is least frequent for the longest window size, rending it the most desirable choice.  
(For the shortest window $T^+ = 0.87$ (green curve) the abrupt change in the state is small relative to the overall errors, but is nonetheless present and most frequent.)

An assessment of the estimation accuracy at $t^+ = 174$ is provided in panel $b$.
As anticipated, the estimation error grows with distance from the observed layer at $y^+=50$, with the rate of growth being more pronounced for shorter window sizes.
These trends can be understood through the lens of the adjoint-variational algorithm. The sensitivity of the outer observations to the near-wall velocity diminishes as we move away from the observations towards the wall. When the window size is short, the region of sensitivity reduces, because a change in the near-wall region does not have time to impact the observation locations.
Based on these results, we recommend using longest possible $T$, which is set to be on the order of the Lyapunov timescale. This choice maximizes the estimation accuracy and minimizes the number of discontinuities of the trajectories in state space.

\section{Linear stochastic estimation of the wall layer}
\label{sec:LSE}
In this appendix, we briefly introduce linear stochastic estimation \citep{Adrian1988LSE,Baars2016}, which is used to construct the first guess of the initial condition in the unknown wall layer $y \in [0,l_y)$.
Although observations are available throughout the outer layer $y \in [l_y,2-l_y]$, only the planar data at $y = l_y$ are used in the linear stochastic estimator.
Including more observations from $y > l_y$ does not improve the estimation accuracy, because these observations are less correlated with the wall layer than the velocity data at $y = l_y$.

The linear stochastic estimation requires knowledge of the covariance matrix between the measurements to the velocities that are to be estimated.  In the present work, this covariance was evaluated using a direct numerical simulation which, importantly, is independent from the true flow that generated the observations for the data assimilation study reported in the main text.  Specifically, the velocity field that was used for computing the LSE is denoted $\boldsymbol{u}_A$ and the associated measurements are $\boldsymbol{m}$.  Once the LSE is constructed, it provides a contemporaneous estimate of the near-wall velocity $\boldsymbol{u}_e$ from available measurements.

To simplify the derivation, we start with fully-resolved outer velocity observations at $y = l_y$, which are denoted $m_j(x,l_y,z,t)$ with $j = \{1,2,3\}$ for the three velocity components.
The linear stochastic estimation of contemporaneous near-wall streamwise velocity is constructed as,
\begin{equation}
    \label{eq:LSE}
    u_e(x,y,z,t) = \int_0^{L_x}\int_0^{L_z} A_{j}(x - x^{\prime},y,z-z^{\prime}) m_j(x^{\prime},l_y,z^{\prime},t)\mathrm{d}x^{\prime} \mathrm{d}z^{\prime}.
\end{equation} 
Here, the repeated indices imply summation.
The linear stochastic estimator $A_{j}$ is assumed to be independent of time.
Furthermore, due to the homogeneity in $x$ and $z$, the estimator $A_{j}$ depends only on the separations between the observation sites and estimation points along the horizontal directions.
Computation of $A_{j}$ is more efficient in Fourier space and requires the knowledge of the second-order statistics of the turbulence.  We proceed with the derivation of $A_{j}$ in Fourier space.

Since equation (\ref{eq:LSE}) is essentially a convolution of $A_{j}$ and $m_j$, applying the Fourier transform with respect to $x$ and $z$ on both sides yields
\begin{equation}
    \label{eq:LSE_Fourier}
    \hat{u}_e(k_x,y,k_z,t) = \hat{A}_{j}(k_x,y,k_z) \hat{m}_j(k_x,l_y,k_z,t),
\end{equation}
which shows that the estimated velocity at wavenumber $(k_x,k_z)$ is a product between the estimator and the observation at the corresponding wavenumber.
The linear estimator $\hat{A}_{j}$ is determined by minimizing the mean-square error between the estimated $\hat{u}_e$ and the true near-wall velocity $\hat{u}_A$:
\begin{equation}
    \label{eq:LSE_loss}
    \left\langle| \hat{u}_e - \hat{u}_A |^2 \right\rangle_t = \left\langle| \hat{A}_{j}(k_x,y,k_z) \hat{m}_j(k_x,l_y,k_z,t)- \hat{u}_A(k_x,y,k_z,t)  |^2 \right\rangle_t.
\end{equation}
The optimal estimator that minimizes equation (\ref{eq:LSE_loss}) satisfies the linear system
\begin{equation}
    \label{eq:LSE_sol}
    \left\langle \hat{m}_i^*(k_x,l_y,k_z,t) \hat{m}_j(k_x,l_y,k_z,t)\right\rangle_t \hat{A}_{j}(k_x,y,k_z) = \left\langle \hat{m}_i^*(k_x,l_y,k_z,t) \hat{u}_A(k_x,l_y,k_z,t)\right\rangle_t,
\end{equation}
where $\langle\hat{m}_i^* \hat{m}_j\rangle_t$ is the covariance matrix of the outer observations at $(k_x,l_y,k_z)$, and $\langle\hat{m}_i^* \hat{u}_A\rangle_t$ is the covariance between the observation and near-wall velocity.
Equation (\ref{eq:LSE_sol}) is solved for each $(k_x,y,k_z)$ respectively to obtain the linear stochastic estimator.

In summary, computing the linear stochastic estimator requires the following three steps. First, we model or compute the covariance matrices in equation (\ref{eq:LSE_sol}).
The covariance matrices are then used to solve for the spectral estimator $\hat{A}_j(k_x,y,k_z)$ using (\ref{eq:LSE_sol}).
Second, given the outer-flow observations at the initial time $t = 0$, we apply the linear estimator (\ref{eq:LSE_Fourier}) to obtain the estimated velocity field at each wavenumber and wall-normal location in the unknown layer, $(k_x,y,k_z)$.
For cases where the outer observations are not fully resolved, equation (\ref{eq:LSE_Fourier}) is applied to wavenumbers within the measurement resolution only, i.e., $k_x \leq k_{x,m}$ and $k_z \leq k_{z,m}$.
Finally, the estimated velocity field is transformed from Fourier space back to physical space to obtain the initial condition in the wall layer.

\bibliographystyle{jfm}
\bibliography{reference}

\end{document}